\documentclass[prd,showpacs,twocolumn,superscriptaddress,preprintnumbers,amsmath,amssymb,nofootinbib]{revtex4-1}
\usepackage{latexsym}
\usepackage{amsmath}
\usepackage{amsfonts}
\usepackage{amssymb}
\usepackage{textcomp}
\usepackage{epsf}
\usepackage{cancel}
\usepackage[normalem]{ulem}
\usepackage{graphicx,color}

\def\XXint#1#2#3{{\setbox0=\hbox{$#1{#2#3}{\int}$}
     \vcenter{\hbox{$#2#3$}}\kern-.5\wd0}}

\def\chpt{$\chi$PT}
\def\rcht{R$\chi$T}
\def\cO{{\cal O}}
\def\1{\'{\i}}
\def\no{\nonumber}

\begin{document}

\title{Large-$N_c$ naturalness in coupled-channel meson-meson
  scattering}
\author{T. Ledwig}
\affiliation{Departamento de F\1sica Te\'orica, IFIC,
Universitat de Val\`encia -- CSIC, Apt. Correus 22085, E-46071 Val\`encia, Spain}

\author{J. Nieves}
\affiliation{Departamento de F\1sica Te\'orica, IFIC,
Universitat de Val\`encia -- CSIC, Apt. Correus 22085, E-46071 Val\`encia, Spain}

\author{A. Pich}
\affiliation{Departamento de F\1sica Te\'orica, IFIC,
Universitat de Val\`encia -- CSIC, Apt. Correus 22085, E-46071 Val\`encia, Spain}

\author{E. Ruiz Arriola}
\affiliation{Departamento
  de F\'isica At\'omica, Molecular y Nuclear and Instituto Carlos I de
  Fisica Te\'orica y Computacional, \\ Universidad de Granada, E-18071
  Granada, Spain}

\author{J. Ruiz de Elvira}
\affiliation{Helmholtz-Institut f\"ur Strahlen- und Kernphysik (Theorie) and
Bethe Center for Theoretical Physics, Universit\"at Bonn }

\pacs{12.38.Gc, 12.39.Fe, 14.20.Dh}

\begin{abstract}
The analysis of hadronic interactions with effective field theory
techniques is complicated by the appearance of a large number of
low-energy constants, which are usually fitted to data. On the
other hand, the large-$N_c$ limit helps to impose natural
short-distance constraints on these low-energy constants, providing
a parameter reduction.  A Bayesian interpretation of the expected
$1/N_c$ accuracy allows for an easy and efficient implementation of
these constraints, using an augmented $\chi^2$.  We apply this
approach to the analysis of meson-meson scattering, in conjunction
with chiral perturbation theory to one loop and coupled-channel
unitarity, and show that it helps to largely reduce the many existing
ambiguities and simultaneously provide an acceptable description of
the available phase shifts.
\end{abstract}

\maketitle

\section{Introduction}

While the solution of QCD remains a difficult and challenging problem
which is being progressively tackled on the lattice, there are two
limits where substantial simplifications apply in the continuum: the
chiral limit~\cite{Langacker:1973hh}, where the current quark mass
$m_q$ is set to zero (see \cite{Pich:1995bw,Ecker:1994gg} for a
review), and the limit of a large number of colours
$N_c$~\cite{'tHooft:1973jz,Witten:1979kh} (see
Ref.~\cite{Lucini:2013qja} for a recent review and references
therein), where the strong coupling constant scales as $\alpha_s \sim
1/N_c$.  The main common virtue of these simplifications is that at
sufficiently low energies, $\sqrt{s} \le \Lambda $, quark-hadron
duality and confinement require that these limits and their
corrections can be expressed in purely hadronic terms, with no
explicit reference to the underlying quark and gluon degrees of
freedom.  A well-known example of this duality is given by the
Gell-Mann--Oakes--Renner relation, $ 2 m_q |\langle \bar q q \rangle| =
f_\pi^2 m_\pi^2 $, which is ${\cal O} ( N_c m_\pi^2 )$ and relates
the current quark mass $m_q$ and the quark condensate $\langle \bar q
q \rangle $ with the pion decay constant $f_\pi$ and the pion mass
$m_\pi$.

Of course, none of these extreme limits are generally expected to
faithfully feature the real world. Instead, the {\it smallness} of the
quark mass as compared to $\Lambda_{\rm QCD}$ and the {\it largeness}
of $N_c=3 $ as compared to unity suggest a sensible hierarchy where an
expansion in the $u,d,s$ quark masses and a $1/N_c$ expansion may be
combined in a suitable way to attempt a credible description of hadron
properties and their interactions.  Within an effective Lagrangian
approach~\cite{Weinberg:1978kz}, and using the low-energy degrees of
freedom (Goldstone bosons) of the non-linear sigma
model~\cite{Appelquist:1980ae}, a chiral perturbation theory (\chpt)
to one loop was thus designed~\cite{Gasser:1983yg,Gasser:1984gg}. On
the other hand, the leading tree-level structure implied by the
large-$N_c$ limit suggests using a resonance chiral theory (\rcht) to
successfully saturate the low-energy
properties~\cite{Ecker:1988te,Ecker:1989yg,Cirigliano:2006hb}.  It has
been shown that resonance saturation arises quite naturally
\cite{Pich:2002xy} from the short-distance constraints on the
effective hadronic theory stemming from the underlying high-energy
behaviour of QCD for space-like momenta.

This scheme is implemented in terms of chiral effective Lagrangians
displaying explicitly the relevant hadronic degrees of freedom,
characterized by i) a {\it finite} number of fields representing
stable particles, in the large-$N_c$ limit, with masses $M_R = {\cal
  O} (N_c^0) \lesssim \Lambda $ , ii) ${\cal O} (m_q)$ suppressed
couplings to pseudoscalar-mesons  and iii) $n-$mesonic ${\cal O}
(N_c^{1-n/2})$ suppressed couplings. The decay rate of these states is
suppressed, $\Gamma_R = {\cal O} (N_c^ {-1})$, and are thus
resonances. The calculation of quantum corrections, besides restoring
unitarity perturbatively within the relevant $\Lambda$-truncated
Hilbert space, accounts for the scale dependence of the couplings in
the Lagrangian, as they effectively and implicitly incorporate the
degrees of freedom which have been integrated out. The number of
low-energy couplings (LECs) depends on how many independent terms can
be written in the effective Lagrangian with fields and their
derivatives, so that they naturally scale with inverse powers of the
breakdown scale $\Lambda$. As it is well known, this number grows
rapidly with the order of the expansion, and predictive power relies
heavily on having more data than couplings. Large-$N_c$ arguments have
helped in fixing the bulk of the scale-independent contribution for
the LECs at ${\cal O}(p^4)$ and ${\cal O}(p^6)$ (see
however~\cite{Kampf:2006bn} for an exception at ${\cal O}(p^6)$).

In this paper we are concerned with the implications of
next-to-leading order (NLO) \chpt\ and leading order (LO) $1/N_c$
corrections in the description of the interactions among pseudoscalar
mesons belonging to the flavour octet and below a given energy
cut-off, which will be set at $\sqrt{s_{\rm max}} \sim
1.1-1.2~\mathrm{GeV} \equiv \Lambda_R $ for definiteness. This energy
cut-off provides a motivation to truncate the infinite tower of meson
states to just one per quantum number (except for the $0^{++}$ scalar
and $0^{-+}$ pseudo-scalar multiplets, where independent octet and
singlet states are considered).
At first order in the $1/N_c$ expansion, terms with more than one trace and loops are
suppressed; therefore, we only include tree-level resonance
contributions. Moreover, we neglect interactions between different resonance channels which, 
although allowed by the theory symmetries, are not needed in this work.

Furthermore, as argued in Ref.~\cite{Nieves:2011gb},
in addition to the tree-level meson-exchange diagrams, one
should also foresee contact pieces, which depend on the high-energy
cut-off $\Lambda_R$.  In that work, $\Lambda_R \sim 700~\mathrm{MeV}$
and thus only elastic $\pi\pi$ scattering was possible. On the other
hand, when $\Lambda_R \sim 1.2~\mathrm{GeV}$ all
pseudoscalar-pseudoscalar channels are open; thus, coupled-channel
unitarity plays a decisive role.  We use here the Bethe-Salpeter
equation (BSE) approach, within the on-shell renormalization
scheme~\cite{Nieves:1999bx} conveniently extended to the coupled-channel case,
which restores exact two-body unitarity, thereby enlarging the scope
of the previous work~\cite{Nieves:2011gb} to include coupled
channels. This BSE on-shell scheme is characterized by the appearance
of non-perturbative subtraction constants, $C_{IJ}$, and a
perturbative matching procedure to reproduce LO $1/N_c$ and NLO \chpt.
In all, we need 24 independent parameters which must be fixed from
fitting scattering data or pseudo-data, a rather impractical
situation. We will show how a judicious fitting strategy, based on the
natural expectation that NLO $1/N_c$ corrections are at the $\sim
30\%$ level, provides good fits with reasonable parameters.

Before embarking into a more involved discussion, let us explain
the main idea behind the present work. In the chiral limit, $m_\pi
\to 0$, QCD has only one dimensionful parameter which can be
chosen to be the pion weak decay constant, $f$.
In the large-$N_c$ limit $f= {\cal O} (\sqrt{N_c})$
and hence meson masses must scale as $M_R \sim f /\sqrt{N_c}$ and
meson couplings as $G_R \sim f $. That means that we expect an
expansion of the form,
\begin{eqnarray}
  \frac{\sqrt{N_c} M_R}{f } = a_R \left[ 1 + \frac{\xi_R}{N_c} + \dots \right]
\label{eq:nc-expansion}
\end{eqnarray}
where $a_R$ and $\xi_R$ are numerical dimensionless coefficients of
order one. The basic idea to be explored in the present work is the use of
this information when we have a good estimate for the LO $a_R$, but no
information on the NLO term $\xi_R$. Under these circumstances, we may
{\it assume} that $\xi_R$ is a random variable normally distributed,
$\langle \xi_R \rangle =0$ and $\langle \xi_R^2 \rangle ={\cal O}(1)$, in which
case the variable (for simplicity couplings $G_R$ are omitted)
\begin{eqnarray}
\chi^2_{\rm th} =\sum_R \xi_R^2 \, , \qquad  \xi_R = \frac{\sqrt{N_c}
  M_R/a_R-f}{f/N_c}
\label{eq:chi2-in}
\end{eqnarray}
follows a $\chi^2$ distribution if the different $\xi_R$'s are
uncorrelated, $\langle \xi_R \xi_R' \rangle = \delta_{R,R'} \langle \xi_R^2 \rangle$. In the
absence of further information, $\chi^2$ is minimized by our guess of
the parameters $M_R$ and $G_R$.  However, if we have further data
which can be described theoretically by these parameters, $M_R, G_R$,
we may refine our initial guess
by adding
to Eq.~(\ref{eq:chi2-in}) the standard $\chi^2$ term used to carry out a fit to these data.
In the first part of the paper we will analyze how the LO coefficients
can be estimated. In the second part we will show how to profit from
these estimates in meson-meson scattering.

The paper is organized as follows.  In Section~\ref{sec:sdc} we
discuss and motivate short-distance constraints in the large-$N_c$
limit and their consequences in the light of data and lattice studies.
In Section~\ref{sec:natural} we analyze how the naturalness of leading
$1/N_c$ corrections provides a sensible estimate on the expected
uncertainties of low-energy constants. The formalism for
coupled-channel unitarized meson-meson scattering and some specific
features are reviewed in Section~\ref{sec:mm-scat}. Our fitting
strategies and main numerical results are presented in
Section~\ref{sec:fit}. Finally, in Section \ref{sec:concl} we
summarize our points and come to the conclusions.

\section{Large-$N_c$ short-distance constraints}
\label{sec:sdc}

\subsection{Motivation}

In this section we analyze some important conditions which arise
from imposing the best possible high-energy behaviour of field
correlators in a low-energy truncated hadronic theory, compatible
with the known behaviour in QCD. This leads to a sensible parameter
reduction, based on estimates of $a_R$ in Eq.~\eqref{eq:nc-expansion}, which will be very helpful
in our analysis of meson-meson scattering. These short-distance
constraints are obtained within a large-$N_c$ framework and when the
low-energy theory is limited to spin 0 and 1 resonances, and allow
for a complete parameter reduction in the chiral limit; all
couplings and masses can be explicitly expressed in terms of the
pion weak decay constant and the number of colours $N_c$. In order
to appreciate the result, it is important to spell out which are the
main assumptions and approximations leading to it.

As already mentioned, in the large-$N_c$ limit interactions among
hadrons and with external currents are suppressed. This allows to set
up a hierarchy in terms of an infinite number of quantum hadronic
fields $R_i(x)$ and their derivatives, compatible with the symmetries
at all energies, which can be written at tree level in terms of 
a real local Lagrangian with a given set of coupling constants $G_{R_i}$ and
hadron masses $M_{R_i}$. Unitarity is recovered perturbatively by
computing quantum corrections in a loop expansion.  As we are
interested in an intermediate energy description, say $\sqrt{s} \le
\Lambda_R$, we need only to consider explicitly a finite number of fields
which are active below this cut-off scale, $M_{R_i} \lesssim
\Lambda_R$. Heavier states, $\Lambda_{R} \lesssim M_{R_i}$ are included
implicitly through the couplings $G_{R_i}
(\Lambda, M_{R_i})$ and masses $M_{R_i} (\Lambda, M_{R_i})$ appearing
in the low-energy Lagrangian.
In our case we will take $\Lambda_R \sim 1.1~{\rm GeV}$. This means
in practice taking, besides the pion, explicit
$S(0^{++})$, $P(0^{-+})$, $V(1^{--})$ and $A(1^{++})$ fields whose masses are
smaller than the cut-off, $m_V,m_S,m_P,m_A \lesssim \Lambda_R$.

\subsection{Resonance Lagrangian}

The \rcht\ Lagrangian describes the dynamics of Goldstone and massive
meson multiplets of the type $S(0^{++})$, $P(0^{-+})$, $V(1^{--})$ and
$A(1^{++})$~\cite{Ecker:1988te,Ecker:1989yg,Cirigliano:2006hb}, in
terms of a set of masses $m_\pi,m_S,m_P,m_V,m_A$ and couplings
$F_A,F_V,G_V,c_d,c_m,d_m, \tilde c_d,\tilde c_m,\tilde d_m,f_\pi$,
which can be determined phenomenologically.  
We will only need the lowest-order couplings:
\begin{eqnarray}
{\cal L}_2 &=& {f^2\over 4}\,\langle D_\mu U^\dagger D^\mu U \, + \, U^\dagger\chi  \,
+  \,\chi^\dagger U\rangle
\no\\ 
&+&  \sum_i\;\left\{ {F_{V_i}\over 2\sqrt{2}}\;
   \langle V_i^{\mu\nu} f_{+ \, \mu\nu}\rangle\, +\,
   {i\, G_{V_i}\over \sqrt{2}} \,\,\langle V_i^{\mu\nu} u_\mu u_\nu\rangle
   \right\}
\no\\
& + & \sum_i\; {F_{A_i}\over 2\sqrt{2}} \;
   \langle A_i^{\mu\nu} f_{- \, \mu\nu} \rangle
   \, +\,\sum_i\; i\, d_{m_i}\;\langle P_i\, \chi_- \rangle
\no\\
& + & \sum_i\;\biggl\{ c_{d_i} \; \langle S_i\, u^\mu
u_\mu\rangle\, +\, c_{m_i} \; \langle S_i\, \chi_+ \rangle\biggr\}\, .
\end{eqnarray}
The matrix $U(\phi) =  u(\phi)^2 = \exp{\left\{i\sqrt{2}\Phi/f\right\}}$ contains
the Goldstone fields, $\chi = 2 B_0 {\cal M}$ is the explicit breaking of chiral symmetry through the quark mass matrix ${\cal M}$,
$u_\mu \equiv i\, u^\dagger D_\mu U u^\dagger$, \
$f^{\mu\nu}_\pm\equiv u F_L^{\mu\nu} u^\dagger\pm  u^\dagger F_R^{\mu\nu} u$,
$\chi_\pm\equiv u^\dagger\chi u^\dagger\pm u\chi^\dagger u$
\ and \ $\langle\,\rangle$ denotes a 3-dimensional flavour trace.
We refer to Refs.~\cite{Ecker:1988te,Ecker:1989yg,Cirigliano:2006hb} for notations and technical details.

Clearly, in the chiral
limit all dimensionful quantities should scale with $\Lambda_{\rm
  QCD}$ or, alternatively, with $f\approx f_\pi \sim \Lambda_{\rm QCD}$. As we
now discuss, it is remarkable that a combination of the large-$N_c$
limit with a set of short-distance constraints, based on imposing
asymptotic QCD conditions stemming from the operator product expansion
(OPE), and with a minimal hadronic ansatz, yields quite naturally to
this scaling behaviour.

\subsection{Short distance constraints at leading order}

Two-, three- and four-point-function constraints have been discussed in
Ref.~\cite{Pich:2002xy} for $SS$, $VV$, $AA$, $PP$, $VPP$ and $SPP$
correlators.  They determine the \rcht\ couplings in terms of the pion
decay constant. For illustration purposes, we review here some of
the short-distance constraints discussed in~\cite{Pich:2002xy}.  At
leading order in $1/N_c$, 
the two-Goldstone matrix element of the vector current and the 
matrix element of the axial current between one Goldstone and one photon are characterized, respectively, by the vector and axial-vector form factors:
\begin{eqnarray}
F_V(t)&\, =&\, 1+\sum_i{\frac{F_{V_i}G_{V_i}}{f_\pi^2}}\frac{t}{m_{V_i}^2-t}\, ,\nonumber\\
G_A(t) &\, =&\, \sum_i{\frac{2F_{V_i}G_{V_i}-F_{V_i}^2}{m_{V_i}^2}+\frac{F_{A_i}^2}{m_{A_i}^2-t}}\, .
\end{eqnarray}
Since they should vanish at $t\rightarrow \infty$,
the resonance couplings should satisfy:
\begin{eqnarray}\label{VVAA}
&&\sum_i{F_{V_i}G_{V_i}}\; =\; f_\pi^2\, ,\nonumber\\
&&\sum_i{\frac{2F_{V_i}G_{V_i}-F_{V_i}^2}{m_{V_i}^2}} \; =\;  0\, .
\end{eqnarray}

In the same way, the leading $1/N_c$ contribution to the $\pi K$ scalar form factor
is given by:
\begin{align}
F^S_{K\pi}(t)\; =\; 1+\sum_i\, &\frac{4c_{m_i}}{f_\pi^2}\left[c_{d_i}+
(c_{m_i}-c_{d_i})\frac{m_K^2-m_\pi^2}{m_{S_i}^2}\right]
\nonumber\\ &\times
\frac{t}{m_{S_i}^2-t}\, .
\end{align}
Imposing again that $F^S_{K\pi}(t)$ should vanish when $t\to \infty$, we get:
\begin{eqnarray}\label{SS}
4\sum_i{c_{d_i}c_{m_i}}=f_{\pi}^2\, , \qquad\sum_i{\frac{c_{m_i}}{m_{S_i}^2}(c_{m_i}-c_{d_i})=0}\, .
\end{eqnarray}
Further constraints arising from the Weinberg and $SS-PP$ sum rules
are also discussed in~\cite{Pich:2002xy}. All these large-$N_c$ constraints involve an infinite tower of resonances.

\subsection{Results with single-resonance saturation}

Assuming an exact U(3) symmetry and that, at low energies,
each infinite resonance sum is dominated
by the first meson nonet with the corresponding quantum numbers,
the short-distance constraints
determine the \rcht\ couplings~\cite{Pich:2002xy},

\begin{eqnarray}
  f_\pi & =& F_A = F_V/\sqrt{2} = \sqrt{2}\, G_V
  = 2\, c_d = 2\, c_m = 2 \sqrt{2}\, d_m
  \nonumber \\ &=&
  2 \sqrt{3}\, \tilde c_d = 2 \sqrt{3}\, \tilde c_m = 2 \sqrt{6}\, \tilde d_m\, ,
  \label{eq:chiRT1}
\end{eqnarray}
and give the mass relations $m_A = \sqrt{2}\, m_V$ and $m_P =
\sqrt{2}\, m_S$.  Imposing in addition a proper short-distance
behaviour for the elastic $\pi\pi$ scattering amplitude, it was found
in Ref.~\cite{Nieves:2011gb} that, in the absence of tensor couplings,
$m_S= m_V$. The absolute mass scale can be further related to $f_\pi$
by requiring the $PVV$ form factor to fall at large momentum as
predicted by QCD~\cite{Masjuan:2012sk}.  One finds:\footnote{The
  same relation between $m_V$ and $f_\pi$ was obtained identifying
  the quark-model one-loop pion radius~\cite{Tarrach:1979ta} to
  vector meson dominance~\cite{Bramon:1981sw}.
  This is equivalent to identify the
  resonance-saturation prediction for the \chpt\ LECs with the
  results obtained in the chiral quark model
  \cite{Pich:1995bw,Espriu:1989ff}. 
  Invoking quark-hadron duality, one also obtains the relations for the other
  masses within the spectral quark model~\cite{RuizArriola:2003bs,Megias:2004uj}.}
\begin{eqnarray}
  m_S= m_V = \frac{m_P}{\sqrt{2}}= \frac{m_A}{\sqrt{2}}= \pi
  \sqrt{\frac{24}{N_c}} f_\pi\, .
  \label{eq:chiRT2}
\end{eqnarray}
The first relation complies with $m_S-m_V = {\cal O}(N_c^ {-1})+
{\cal O}(m_\pi N_c^0)$, discussed in Ref.~\cite{Nieves:2009ez} and
obtained after identifying $m_V$ and $m_\rho$ in the large-$N_c$
limit.  This amounts in particular also to the width/mass
ratios\footnote{The first relation is a direct consequence of the
  discussion in Sec. V of Ref.~\cite{Nieves:2009ez} and
  Eq.~(\ref{eq:chiRT2}), while the scalar one is deduced from the
  constraint $\Gamma_S/m_S= 9\Gamma_V/2 m_V$ re-derived, for
  instance, in Ref.~\cite{Nieves:2011gb} in the absence of tensor
  couplings from the Adler and $\sigma$ sum rules within the
  single-resonance approximation scheme.}
\begin{eqnarray}
  \frac{\Gamma_V}{m_V}&=& \frac{\pi}{4 N_c} \left[ 1 + {\cal O} (N_c^{-1}) \right],
  \nonumber \\
  \frac{\Gamma_S}{m_S}&=& \frac{9\pi}{8 N_c} \left[ 1 + {\cal O} (N_c^{-1}) \right],
\end{eqnarray}
which compare rather well with the experimental Breit-Wigner values
for $\Gamma_\rho/m_\rho$ and
$\Gamma_\sigma/m_\sigma$~\cite{Nieves:2011gb}.  As noted in
~\cite{Nieves:2009kh}, the location of the Breit-Wigner and pole
masses differ by ${\cal O}(N_c^{-2})$ corrections.

Although consistent with the large-$N_c$ counting, these relations
assume that the high-energy properties can be properly saturated with
a minimal set of resonances. Therefore, they are subject to corrections
already at LO in $1/N_c$, due to the neglected higher-energy
states. These corrections are difficult to estimate when there are
more massive states than constraints. Actually, in the opposite case,
and for the single-resonance case, there may appear contradicting
constraints~\cite{Bijnens:2003rc} (see the comprehensive discussion in
Ref.~\cite{Roig:2013baa}) which provide similar relations with {\it
  different} and $N_c$-independent numerical factors. There is of
course the pertinent question on {\it what} numerical values should be
used in the $\Lambda_R$-truncated \rcht\ effective Lagrangian, since it
is itself of LO in the $1/N_c$ expansion.

We remind in this regard that resonances manifest as poles of
scattering amplitudes in the second Riemann sheet (SRS), $\sqrt{s_R}=
m_R - i\,\Gamma_R/2$, and in principle have vanishing widths in the
large-$N_c$ limit.\footnote{Despite behaviours of the poles with
  non-vanishing widths in the large-$N_c$ limit have been described in the
  literature~\cite{Nieves:2011gb,
    Pelaez:2003dy,RuizdeElvira:2010cs,Guo:2011pa,Guo:2012yt,Guo:2012ym},
  it has been recently reviewed in~\cite{Cohen:2014vta} that it is not
  possible to find any meson configurations in terms of quark and
  gluons with non-vanishing widths in the large-$N_c$ limit coupled to
  meson-meson channels.} The NLO corrections are ${\cal O} (N_c^
{-1})$, corresponding to a mass shift $\Delta m_R$ and the width
$\Gamma_R$ which are numerically alike~\cite{Masjuan:2012sk}.
Ultimately, QCD determines the proper numerical factors.

\subsection{Comparison with large-$N_c$ lattice calculations}

The large-$N_c$ limit has recently been implemented on the lattice by
numerically changing $N_c=2,3,4,5,6,$
$7\dots$~\cite{Bali:2013kia,Bali:2013fya} and extracting meson masses
and decay widths in the quenched approximation, since corrections are
$1/N_c^2$ suppressed (the fermion determinant providing the LO
corrections in $1/N_c$ was not included). They find $m_\rho/f_\pi =
[7.08\, (10),\, 7.21\, (10)] $, $m_{a_1}/f_\pi = [13.16\, (21),$$
  13.26\, (21)]$, $m_{\pi^*}/f_\pi = [15.61\, (34),\, 15.70\, (34)]$
and $f_\rho/f_\pi= [1.861\, (30), 1.875$ $ (31)]$ for $m_q= [0,\,
  m_{ud}]$, respectively, to be compared with Eq.~(\ref{eq:chiRT2})
where one has $m_V/f_\pi=8.89$, $m_A/f_\pi=12.57$, $m_P/f_\pi=12.57$
and $f_V/f_\pi = 1.41$ (note a $\sqrt{2}$ factor of difference between
the normalization for $f_V$ used here and that of
Ref.~\cite{Bali:2013kia}; regarding to the decay constants, due to the
lack of non-perturbative renormalization at $N_c = \infty$, an error
of 8\% should be associated to the values quoted above and taken from
the Table 4 of Ref.~\cite{Bali:2013kia}).  Unfortunately, no
predictions have yet been made for the troublesome $0^{++}$ scalar
mesons on the lattice at large $N_c$. Nonetheless, the comparison is
good enough to discard a purely accidental agreement with
Eqs.~(\ref{eq:chiRT1}) and (\ref{eq:chiRT2}), not only at the
phenomenological level, but also in the large-$N_c$ limit of QCD.

On dimensional grounds the parameter reduction in QCD, in the
chiral limit, is obvious from a large-$N_c$ counting point of view
and the existence of a unique dimensionful scale $f_\pi$. One
could, of course, take these exact lattice values as an initial
guess for our analysis below; they are subjected to ${\cal
  O}(1/N_c)$ and chiral corrections.  Unlike our estimates, they get
no corrections from higher-energy states. Unfortunately, some of the
needed parameters are still missing, so we will content ourselves
with our estimates for couplings and mass relations,
Eq.~(\ref{eq:chiRT1}) and Eq.~(\ref{eq:chiRT2}) respectively, based
on just one single resonance saturation.

\section{Natural size of $1/N_c$ corrections}
\label{sec:natural}

The structure of the chiral expansion in powers of $m_\pi^2/(4 \pi
f_\pi)^2$ (up to chiral logarithms $\sim \log m_\pi^2$) is well
understood and has been worked out in much detail for many
processes. Actually, the chiral-loop ${\cal O}(p^n)$ corrections are
themselves of ${\cal O}(N_c^{-n/2})$.  However, much less is known on
what are the expected and genuine corrections within a $1/N_c$
expansion. While at LO only tree-level diagrams contribute, with the
exchange of an infinite number of meson resonances, quantum loops with
massive states propagating in the internal lines need to be considered
at the NLO. Sub-leading $1/N_c$ quantum corrections involving a
limited number of resonances have been already
investigated~\cite{Cata:2001nz,Rosell:2004mn,Rosell:2005ai,Rosell:2006dt,Pich:2008jm,
  Pich:2010sm,SanzCillero:2009ap}. A simple rule of thumb is that they
are naturally expected to have a $30\%$ accuracy.  One vivid
demonstration of this naive expectation is given by the width/mass
ratio for meson and baryon resonances which scales as $\Gamma/M \le
{\cal O}(N_c^{-1})$~\cite{'tHooft:1973jz,Witten:1979kh}, suggesting a
relative $30\%$ ratio, whereas the PDG~\cite{Beringer:1900zz}
(spin-weighted) average values {\it both} for mesons and baryons
containing $u,d,s$ flavours are identical and equal to $\langle
\Gamma/M \rangle=0.12\, (8)$~\cite{Arriola:2011en,Masjuan:2012gc},
where the uncertainty itself is compatible with a sub-leading $1/N_c$
correction.

In order to motivate our approach to meson-meson scattering below, we
will illustrate the size of $1/N_c$ corrections for the LECs
$L_{1,\dots, 10}$. As already noted, \rcht\ predicts their
leading-$N_c$ value~\cite{Ecker:1988te,Ecker:1989yg}, but quite
remarkably no errors on that estimate are ever quoted (besides the
scale dependence which is $1/N_c$ suppressed and is usually taken to
be $\mu=m_\rho$).

Using Eqs.~(\ref{eq:chiRT1})
and (\ref{eq:chiRT2}) (see~\cite{Pich:2002xy}), one obtains the
following set of relations among the LECs:\footnote{When the $\eta_1$
  is integrated out, $L_7$ receives a contribution proportional to
  $1/M_{\eta_1}^2 \sim {\cal O}(N_c^2)$ \cite{Gasser:1984gg}. However,
  the large-$N_c$ counting is no-longer consistent if one takes the
  limit of a heavy $\eta_1$ mass ($N_c$ small) while keeping $m_s$
  small~\cite{Peris:1994dh}.}
\begin{eqnarray}
  2 L_1&=& L_2 =-\frac{L_3}2 = \frac{L_5}2  = \frac{L_8}3  =
  \frac{L_9}4 = -\frac{L_{10}}3
  = \frac{N_c}{192 \pi^2} \, ,
  \nonumber \\
  L_4 &=& L_6=L_7=0 \, ,
  \label{eq:largeNc-LECS}
\end{eqnarray}
which are valid up to corrections of ${\cal O}(N_c^0)$.
A rule-of-thumb estimate for the size of the sub-leading corrections is
$\Delta L_i=L_i/N_c$. The situation is illustrated in
Table~\ref{tab:lecs} where we give the phenomenological values of the LECs from
$\cO(p^4)$ and $\cO(p^6)$ fits, compiled in
Ref.~\cite{Ecker:2007dj}.\footnote{The recent global fits
  of Refs.~\cite{Bijnens:2011tb, Bijnens:2014lea} list many results for $L_{1, \dots, 8} (m_\rho)$.
  See also the recent NLO determinations:  $L_9 (m_\rho)=7.9\, (4)\times
  10^{-3}$~\cite{Pich:2010sm}, $L_{10}(m_\rho)= -4.06\, (39)\times
  10^{-3}$~\cite{GonzalezAlonso:2008rf}
  and $L_{10}(m_\rho)= -3.46\, (32)\times 10^{-3}$~\cite{Boyle:2014pja}.}
The column labeled `\rcht' shows the resonance-exchange predictions,
using input values for $m_V$ and
$m_S$~\cite{Ecker:1988te,Ecker:1989yg,Pich:2002xy}.
Finally, the third column collects the estimations stemming from
the short distance constraints of Eq.~\eqref{eq:largeNc-LECS}, with an error $L_i/N_c$.
The agreement with the $\cO(p^4)$ phenomenological LECs is
somehow deteriorated at $\cO(p^6)$, when NNLO chiral corrections are
taken into account in the fits.  The \chpt\ loop contributions
included at NLO and NNLO are themselves of ${\cal O}(N_c^{-1})$ and
${\cal O}(N_c^{-2})$, respectively, but the \rcht\ predictions refer
to the large-$N_c$ limit and, therefore, are subject to $1/N_c$
corrections.  Once our rule-of-thumb expected error of about 33\% is
considered, the fitted values of the LECs are consistent with the
large-$N_c$ estimates. The differences on the values obtained when
alternative short-distance constraints are invoked are also
comparable~\cite{Roig:2013baa}.
\begin{table}
  \begin{centering}
    \begin{tabular}{ccccc}
      \hline
      \hline
      Parameter & \rcht & Eq.~(\ref{eq:largeNc-LECS})
      & ${\cal O}(p^4)$ & ${\cal O}(p^6)$  \\ \hline
      $ L_1$ & 0.90 & $0.75(25)$ & $0.70(30)$ & $0.43(12)$ \\
      $ L_2$ & 1.80 &  $1.5(5)$  & $1.3(7)$ & $0.73(12)$ \\
      $ L_3$ & $-$4.30 &  $-3.0(1.0)$ & $-4.4 (2.5)$ & $-2.35(37)$  \\
      $ L_4$ & 0.00 & $0.0(3)$ & $-0.3(5)$ & $\sim 0.20$ \\
      $ L_5$ & 2.10 & $3.0(1.0)$ & $1.4(5)$ & $0.97(11)$ \\
      $ L_6$   & 0.00 & $0.0(3)$ & $-0.2(3)$ & $\sim 0.00$ \\
      $ L_7$   & $-$0.30 & $0.0(3)$ & $-0.4(2)$ & $-0.31(14)$ \\
      $ L_8$   & 0.80 & $1.2(4)$  & $0.9(3)$ & $0.60(18)$ \\
      $ L_9$   & 7.10  & $6.0(2.0)$ & $6.9(7)$ & $5.93 (43)$ \\
      $ L_{10}$ & $-$5.40 & $-4.50(1.5)$ & $-5.5(7)$ & $-5.09(47)$ \\
      \hline
      \hline
    \end{tabular}
  \end{centering}
  \caption{Predicted values of the \chpt\ LECs (in units of $10^{-3}$)
    obtained from \rcht ~\cite{Ecker:1988te,Ecker:1989yg},
    i.e. the leading-$1/N_c$ single-resonance approximation,
    and from the short-distance constraints of
    Eq.~\eqref{eq:largeNc-LECS},
    compared with ${\cal O}(p^4)$
    and ${\cal O}(p^6)$ phenomenological determinations at
    $\mu=m_\rho$~\cite{Ecker:2007dj}.\label{tab:lecs} }
\end{table}
The upshot of the previous discussion is that we naturally expect the $1/N_c$ corrections
to the \rcht\ parameters $p_i$ to be of the form
\begin{eqnarray}\label{pis}
  p_i \, =\, p_i^{{\rm Large}-N_c}\; \left(1+ \frac{\xi_{i}}{N_c}\right)\, ,
\end{eqnarray}
where $\xi_{i}$ is of order unity, and could in principle be
calculated.  However, if no complete
information is available, we may for the time being {\it assume} that
$\xi_i$ is a random variable, with $\langle \xi_i \rangle=0$ and
$\langle \xi_i \xi_j \rangle= \delta_{ij}\, \langle \xi_i^2
\rangle$. Of course, one can improve the bias $\langle \xi_i \rangle=0
$ by adding chiral corrections explicitly. The important feature is
that this naturalness assumption will impose rough but {\it a priori}
expectations on the values of the LECs. If $\xi_i$ are Gaussian
parameters, then
\begin{eqnarray}
  \chi^2_{\rm th}\; =\;
  \sum_{i=1}^N \xi_i^2 \; =\;  \sum_{i=1}^{N} \,\left(\frac{p_i-p_i^{{\rm Large}-N_c}}{
    p_i^{{\rm Large}-N_c}/N_c}  \right)^2
  \label{eq:chi2-TH}
\end{eqnarray}
follows a $\chi^2$ distribution. This point of view will be very
helpful below when we analyze coupled-channel meson-meson scattering in
the pseudoscalar sector.

\section{Pion-pion and pion-kaon scattering}
\label{sec:mm-scat}

\begin{figure}
  \begin{center}
    \includegraphics[scale=0.7]{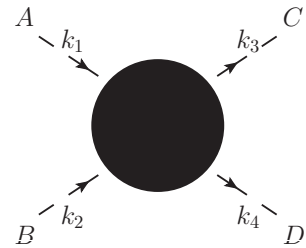}
  \end{center}
  \caption{\label{fig:Meson-meson-scattering.} Meson-meson scattering process $A+B\to C+D$.
    with incoming momenta $k_{1}$, $k_{2}$ and outgoing momenta $k_{3}$, $k_{4}$.
    Dashed lines denote pseudoscalar mesons and the black dot their interaction.}
\end{figure}

We will analyze experimental/phenomenological data for the
$\pi\pi$ and $K\pi$ scattering processes:
\begin{equation}
  \pi\pi \to\pi\pi\, ,
  \qquad
  \pi\pi \to K\overline{K}\, ,
  \qquad
  K \pi \to  K\pi \, .
  \label{eq:piKpiK}
\end{equation}
The two-body kinematics (Fig.~\ref{fig:Meson-meson-scattering.}) is parametrized by the
Mandelstam variables $s=\left(k_{1}+k_{2}\right)^{2}$,
$t=\left(k_{1}-k_{3}\right)^{2}$ and $u=\left(k_{1}-k_{4}\right)^{2}$,
with $\sqrt{s}$ the total energy in the center-of-mass (CM)
system. In our case here, we consider a CM energy ranging from
$\sqrt{s}=280$ MeV up to $\approx 1200$ MeV. At these energies the
following additional channels are open:
\begin{eqnarray}
  &&\pi\pi\to\eta\eta \, ,
  \qquad
  K\overline{K}\to\eta\eta\, ,
  \qquad
  K\overline{K}\to K\overline{K} \, ,
  \nonumber\\
  &&\eta\eta\to\eta\eta\, ,
  \qquad
  K\eta\to K\pi \, .
  \label{eq:KetaKpi1}
\end{eqnarray}
Since in all processes the isospin and strangeness is conserved, one can write each scattering
amplitude in terms of its contributions of total isospin $I$, with
$I=0,\frac{1}{2},1,\frac{3}{2},2$ the only possible values
here, and strangeness.  Choosing $s$ and the scattering angle $\theta$ as the
independent variables, each isospin-projected amplitude
$T_{I}(s)$ can be further decomposed into its individual
contributions with total angular momentum $J$ (for the sake of brevity, we will not make
explicit reference to the strangeness quantum number):
\begin{eqnarray}
  T_{I}(s,t,u) & = & \sum_{J=0}^{\infty}\;\left(2J+1\right)\; T_{IJ}(s)\; P_{J}(\cos\theta)\, ,\label{eq:I-amplitudes}
  \\
  T_{IJ}(s) & = & \frac{1}{2N}\int_{-1}^{+1}d(\cos\theta)\;\, P_{J}(\cos\theta)\; T_{I}(s,t,u)\,
  \nonumber\\
  & = & -16\pi\;\frac{1}{2i\,\rho (s)}\;\left[\eta_{IJ}(s)\; e^{2i\delta_{IJ}(s)}-1\right],
  \label{eq:IJ-amplitudes}
\end{eqnarray}
with $P_{J}(x)$ the Legendre polynomials and
$\rho(s)$ a channel dependent kinematical factor defined
below, and $N$ is a normalization factor to account for identical particles,
such that $N=2$ if all the particles in the process are identical and $N=1$ otherwise.
Since we are working in the isospin limit, we consider the three pions as identical.
Therefore, in our case, $N=2$ only for $\pi\pi\to\pi\pi$ and $\eta\eta\to\eta\eta$ processes.
Explicitly, we analyze data for the three channels in
Eq.~(\ref{eq:piKpiK}) that come in terms of the
scattering amplitudes $T_{IJ}(s)$, phase shifts
$\delta_{IJ}(s)$ and inelasticities
$\eta_{IJ}(s)$, defined in Eq.~(\ref{eq:IJ-amplitudes}).  To
address the resonance properties, a unitarized framework is needed
which leads to the inclusion of coupled channels at higher energies.
Because of this, we need in addition to the above three channels also
a theoretical description of the ones in Eq.~(\ref{eq:KetaKpi1}). In that sense, the present
work is an extension of Ref.~\cite{Nieves:2011gb} to more channels and
higher scattering energies.

Meson-meson scattering within one-loop \chpt\ was analyzed by
G\'omez-Nicola and Pel\'aez in Ref.~\cite{GomezNicola:2001as} where,
in addition, unitarization was implemented via the Inverse Amplitude
Method (IAM). A naive addition of the missing LO contributions in
$1/N_c$, within this scheme, would violate either unitarity or
analyticity.\footnote{The IAM cannot be applied naively to the $1/N_C$
  expansion, because it leads to a re-summation that does not restore
  two-body unitarity. Thus, if $T= T_{1/N_C}+ T_{1/N_C^2} + \cdots$,
  $T_{\rm IAM} = T^2_{1/N_C}/(T_{1/N_C}-T_{1/N^2_C})$ does not fulfill
  the two-body elastic unitarity condition.}  We use here a scheme
based on the Bethe-Salpeter equation (BSE) to restore two-body
unitarity. The BSE on-shell scheme for the non-coupled channel was
already described in Refs.~\cite{Nieves:1999bx,Nieves:1998hp} for
\chpt\ and used in \cite{Nieves:2011gb} when LO $1/N_c$ corrections
were further included. The generalization to the coupled-channels
situation needed here is straightforward. Let us consider the matrix
$T_{IJ}(s)$ incorporating the partial-wave amplitudes of all relevant
processes $AB \to CD$:
\begin{equation}
  T_{IJ}(s)\; =\;\left[\begin{array}{ccc}
      T_{IJ}^{\pi\pi\to\pi\pi}(s) & T_{IJ}^{\pi\pi\to K\overline{K}}(s) & T_{IJ}^{\pi\pi\to\eta\eta}(s)\\
      T_{IJ}^{K\overline{K}\to\pi\pi}(s) & T_{IJ}^{K\overline{K}\to K\overline{K}}(s) & T_{IJ}^{K\overline{K}\to\eta\eta}(s)\\
      T_{IJ}^{\eta\eta\to\pi\pi}(s) & T_{IJ}^{\eta\eta\to K\overline{K}}(s) & T_{IJ}^{\eta\eta\to\eta\eta}(s)
    \end{array}\right]\, ,
  \label{eq:TIJ3x3}
\end{equation}
for the channels $\pi\pi\to\pi\pi$ and $\pi\pi\to K\overline{K}$,
and
\begin{equation}
  T_{IJ}(s)\; =\;\left[\begin{array}{cc}
      T_{IJ}^{K\pi\to K\pi}(s) & T_{IJ}^{K\pi\to K\eta}(s)\\[2pt]
      T_{IJ}^{K\eta\to K\pi}(s) & T_{IJ}^{K\eta\to K\eta}(s)
    \end{array}\right]\, ,
  \label{eq:TIJ2x2}
\end{equation}
for the channel $K\pi\to K\pi$.  All $T_{IJ}^{AB\to CD}(s)$ are
defined through Eq.~(\ref{eq:IJ-amplitudes}), and the explicit form of
the amplitudes $T_{I}\left(s,t,u\right)^{AB\to CD}$ is given in the
Appendix~\ref{sub:APP-Iso-spin-projected-amplitudes}.  Note that some
of the above matrix elements are trivially zero from isospin or
angular momentum conservation.  For instance, in the $I=2$, $J=0$
channel, only $T_{IJ}^{\pi\pi\to\pi\pi}(s)$ is different from zero.

Coupled-channel unitarity is most simply expressed in terms of the inverse matrix $T_{IJ}^{-1}(s)$ as
\begin{eqnarray}
  {\rm Im}  T_{IJ}^{-1}(s)\; =\;
  - {\rm Im} \overline{\mathcal{I}}_{0}(s)\, ,
\end{eqnarray}
where $\overline{\mathcal{I}}_{0}$
is a diagonal matrix of one-loop integrals characterizing the elastic two-body re-scattering:
\begin{equation}
  \overline{\mathcal{I}}_{0}(s)\; =\; \mathrm{diag}\left[
    \overline{I}_{0}^{\pi\pi}(s),\, \overline{I}_{0}^{K\bar K}(s),\,
    \overline{I}_{0}^{\eta\eta}(s)\right]\,
\end{equation}
or
\begin{equation}
  \overline{\mathcal{I}}_{0}(s)\; =\;\mathrm{diag}\left[
    \overline{I}_{0}^{K\pi}(s)\, , \overline{I}_{0}^{K\eta}(s)\right]\
  \label{eq:I0_pipi}
\end{equation}
for the $\pi\pi$ and $K\pi$ cases, respectively.
With two identical mesons,
\begin{equation}
  \overline{\mathcal{I}}_{0}^{\phi\phi}(s)\; =\;
  \frac{1}{16\pi^{2}}\; \rho_{\phi}(s)\;
  \ln\left[\frac{\rho_{\phi}\left(s\right)+1}{\rho_{\phi}\left(s\right)-1}\right]\, ,
\end{equation}
where \ $\rho_{\phi}(s)=\sqrt{1-4m_{\phi}^{2}/s}$ \ and
\begin{equation}
  {\rm Im} \overline{\mathcal{I}}_{0}^{\phi\phi}(s)\; =\;
  -\theta(s-4m_{\phi}^{2})\;\frac{1}{16\pi}\;\rho_{\phi}(s)\, .
\end{equation}
The general expression in the case of two different mesons can be found in
Eq.~(A10) of Ref.~\cite{Nieves:2001wt}, identifying the
function $L(s)$ that appears there  to
${\mathcal{I}}_{0}^{\phi\phi'}(s)/16\pi^2$. The definition/extension
of the loop function to the SRS is given in Eq.~(A13) of the same work.

We decompose the full \chpt\ amplitude matrix $T_{IJ}^{\chi
  PT}\left(s\right)$ in its $\mathcal{O}(p^{2})$ and
$\mathcal{O}(p^{4})$ contributions (in matrix notation):
\begin{eqnarray}
  T_{IJ}^{\chi PT}(s) &\, = &\, T_{IJ}^{\left(2\right)}(s)+T_{IJ}^{\left(4\right)}(s)+\mathcal{O}\left(p^{6}\right).
\end{eqnarray}
The coupled-channels unitarized amplitude is now written as:
\begin{equation}
  T_{IJ}^{-1}\left(s\right)\; =\;
  V_{IJ}^{-1}(s) - \overline{\mathcal{I}}_{0}(s)-\mathcal{C}_{IJ}\, ,
  \label{eq:UACC}
\end{equation}
with $\mathcal{C}_{IJ}$ a diagonal matrix of subtraction constants:
\begin{equation}
  \mathcal{C}_{IJ}\; =\;\mathrm{diag}\left[ C_{IJ}^{\pi\pi},\, C_{IJ}^{K\bar K},\, C_{IJ}^{\eta\eta}\right]
\end{equation}
or
\begin{equation}
  \mathcal{C}_{IJ}\; =\;\mathrm{diag}\left[ C_{IJ}^{K\pi},\, C_{IJ}^{K\eta}\right]\, .
\end{equation}
The matrix $V_{IJ}(s)$ is defined such that a chiral expansion of
$T_{IJ}(s)$ and $V_{IJ}(s)$ will match the one of $T_{IJ}^{\chi
  PT}(s)$. Inverting Eq.~(\ref{eq:UACC}), we obtain the unitarized
matrix $T_{IJ}(s)$:
\begin{eqnarray}
  T_{IJ}(s) & = &
  \left[V_{IJ}^{-1}(s)-\overline{\mathcal{I}}_{0}(s)-\mathcal{C}_{IJ}\right]^{-1}\,
  ,\label{eq:tunita}
  \\
  V_{IJ}(s) & = & T_{IJ}^{\chi PT}(s)-T_{IJ}^{\left(2\right)}
  \left[\overline{\mathcal{I}}_{0}(s)+\mathcal{C}_{IJ}\right]T_{IJ}^{\left(2\right)}.
  \label{eq:vunita}
\end{eqnarray}
\begin{figure}
  \begin{center}
    \includegraphics[scale=0.3]{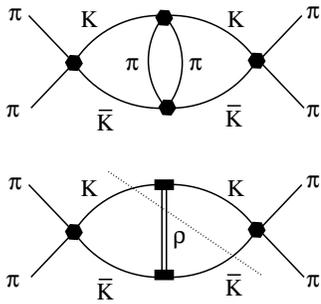}
  \end{center}
  \caption{\label{fig:uni-problems} Some ${\cal O}(p^8)$ loop
    contributions generated when coupled-channels unitarity is restored
    (Eqs.~(\ref{eq:tunita}), (\ref{eq:vunita}) and (\ref{eq:unita-sra})).
    They give rise to pathologies when estimated
    using an on-shell approximation. In the bottom panel, the line shows
    a possible source of a physical imaginary part, when the particles
    cut by it are put on the mass shell.}
\end{figure}

As already noted in Ref.~\cite{GomezNicola:2001as} within the IAM
method, the on-shell approximation in the coupled-channel potential
has the drawback of generating spurious singularities below the
opening of a new channel because the left-cut is analytically
extrapolated below the inelastic threshold.  For instance, as can be
appreciated in the top panel of Fig.~\ref{fig:uni-problems}, for
$\pi\pi \to \pi\pi$ below the  $K\bar K$ threshold, the three loops
$\pi\pi \to \bar K K \to \bar K K \to \pi\pi $ term contains in the
{\it on-shell} $\bar K K \to \bar K K$ piece a $2\pi$-exchange
contribution in the $t$ channel, generating a left-cut in the partial
waves at $s= -4 m_\pi^2+4 m_K^2 $ which sits in the elastic scattering
region $ 4 m_\pi^2 < s < 4 m_K^2 $. However, the effect is numerically
quite small.

\subsection{$\mathcal{O}(p^{4})$ \chpt~ $\&$ single-resonance
  approximation (SRA) amplitudes and large$-N_c$ counting rules}

To incorporate LO $1/N_c$ corrections in the description of the
interactions among the Goldstone bosons, we follow the scheme derived
in \cite{Nieves:2011gb}. There, the leading-$N_c$ prediction for the
actual $\pi\pi$ scattering amplitude was used as deduced from \rcht,
considering just the lowest-lying nonet of
exchanged resonances \cite{Ecker:1988te,Ecker:1989yg}.\footnote{This
  latter approximation is justified as long as $s, t$ and $u$ are kept
  far from the second resonance region.} Thus, let us denote by
$T_{IJ}^{\rm SRA}(s)$, the two SU(3)-Goldstone-boson scattering
amplitude within the SRA, obtained from the lowest-order \rcht\,
Lagrangian~\cite{Ecker:1988te,Ecker:1989yg}, and projected onto
isospin and angular momentum.  Below the resonance mass scale, the
singularity associated with the pole of a resonance propagator could
be replaced by the corresponding momentum expansion; therefore, the
exchange of virtual resonances generates derivative Goldstone
couplings proportional to powers of $ 1/m^2_R$. Let us denote by
$T_{IJ}^{\rm (4)\,SRA}(s)$ the lowest-order term in derivatives. It
gives the large-$N_c$ predictions for the $\mathcal{O}(p^{4})$ \chpt\,
couplings~\cite{Ecker:1988te,Ecker:1989yg} and constitutes the leading
$1/N_c$ approximation to $T_{IJ}^{\chi PT}$. Our approach consists in
using Eqs.~(\ref{eq:tunita}) and (\ref{eq:vunita}), but replacing
in the definition of the two particle irreducible amplitude
$V_{IJ}(s)$, $T_{IJ}^{\chi PT}(s)$  by
\begin{equation}
  T_{IJ}^{\chi PT}(s) \leftrightarrow \underbrace{T_{IJ}^{\rm
      SRA}(s)}_{{\cal O}(1/N_c)}+ \underbrace{\left(T_{IJ}^{\chi
      PT}(s) -T_{IJ}^{\rm(4)\,SRA}(s)\right)}_{{\cal O}(1/N^2_c)}.\label{eq:unita-sra}
\end{equation}
In this way, by construction, we recover the one-loop \chpt\,
results, while at the same time all terms in the amplitude
that scale like $1/N_c$ (leading) are also included within the
SRA. Note that in the $1/N_c$ counting, the correction $\left(T_{IJ}^{\chi
  PT}(s) -T_{IJ}^{\rm(4)\,SRA}(s)\right)$ is incomplete, since it does
not account for all existing subleading $1/N_c^2$ contributions
to $T_{IJ}(s)$. A complete $1/N^2_c$ calculation would require
quantum corrections stemming also from the low-lying
resonances.

We should point out a problem that now appears when unitarity is
restored. Let us pay attention for instance to the $\rho$-exchange,
for $\pi\pi \to \pi\pi$ below the $K\bar K$ threshold. The $\pi\pi \to
\bar K K \to \bar K K \to \pi\pi $ term contains a
contribution from the intermediate $\bar K K \to \bar K K$ amplitude
driven by $\rho$-exchange in the $t-$channel (see bottom panel of
Fig.~\ref{fig:uni-problems}). Such contribution, within the on-shell
scheme used here, leads to a spurious left-cut contribution at $s\le 4
m_K^2-m_\rho^2\sim (0.64 ~{\rm GeV})^2$, with very visible
consequences if nothing is done. It is indeed unphysical, and it is an
artifact of the on-shell unitarization in coupled channels adopted
here. A contribution as the one described above can not physically
give rise to an inelastic imaginary part below $s = (2 m_K+m_\rho)^2$,
as trivially inferred from the optical theorem (see the shown cut in
the figure).  In our case we handle the problem by smoothly switching
off coupled channels effects for $\sqrt{s} \le 0.73$ GeV, and
considering thus purely elastic $\pi\pi\to \pi \pi$ scattering below
these energies, where coupled channel effects are expected to be
negligible. The same procedure has been applied to other channels,
where similar problems also show up. This problem appears but was not
mentioned explicitly in Ref.~\cite{Guo:2011pa} and it was solved there
by re-expanding the $\rho$-meson propagator as a polynomial, hence
removing the singularity.
Note that, the truncation of the expansion implies that in
\cite{Guo:2011pa} not all leading $1/N_c$ terms in the amplitude are
included.

\section{Fitting strategies}
\label{sec:fit}

\subsection{Fitting parameters}

Our 24 fitting parameters can be separated into partial-wave specific
ones which play the role of renormalization constants,
\begin{eqnarray}\label{eq:C's}
  && C_{00}^{\pi\pi}\, ,\, C_{11}^{\pi\pi}\, ,\, C_{20}^{\pi\pi}\, ,\, C_{00}^{K\bar K}\, ,\, C_{00}^{\eta\eta}\, ,\, C_{11}^{K\bar K}\, ,
  \nonumber  \\
  && C_{\frac{1}{2}0}^{K\pi}\, ,\, C_{\frac{1}{2}0}^{K\eta}\, , \, C_{\frac{1}{2}1}^{K\pi}\, ,\, C_{\frac{1}{2}1}^{K\eta}\, ,\, C_{\frac{3}{2}0}^{K\pi}\, ,\, C_{\frac{3}{2}1}^{K\pi}\, ,
\end{eqnarray}
and those which appear in the potential:
\begin{equation}
  G_V\, ,\, c_d \, ,\, c_m \, ,\, \tilde c_d \, ,\, \tilde c_m \, ,\, d_m\, ,\, \tilde d_m \, ,\, m_{S_1}\, ,\, m_{S_8}\, ,\, m_V\, ,\, \mu_V\, ,\, \mu_S\, .
  \label{eq:chiRTpar}
\end{equation}
The \rcht\ predictions are supposed to be valid at some fixed value
of the renormalization scale, which we allow to be different for
vector ($\mu_V$) and scalar ($\mu_S$) couplings.\footnote{The low-lying
vector, axial-vector, scalar and pseudoscalar resonances contribute to
the $L_i$ (see Table~\ref{tab:lecs}), which  renormalized
vales $L^r_i$ can be written as a sum $L^r_i= L_i^{\rm
  SRA}+ \hat L_i^r(\mu)$ of the resonance contributions, and a
remainder $\hat L_i^r(\mu)$. The choice of the renormalization scale
$\mu$ is arbitrary, and it is common to adopt $\mu = m_\rho$ as a
reasonable choice. However, one
might take as a best fit parameter one scale, $\mu^{\rm RS}$,
for which a complete resonance saturation of all the LECs $L^r_i$ occurs
, this is to say $\hat L_i^r(\mu^{\rm RS})=0$. As suggested in
\cite{Nieves:2011gb}, we have  considered a scenario where the
complete resonance saturation of the LECs $L^r_i$
occurs at two different scales, $\mu_V$ for $L^r_{1,2,9,10}$ and $\mu_S$ for
$L^r_{4,5,6,8}$ depending whether the LEC is dominated by
the vector or the scalar resonance contribution. Note
that, $L_3$ and $L_7$ are renormalization-scale invariant.}

\subsection{Fitted data and error assignment}
\label{Sec-Fitted-data-and-error-assignment}
An important novelty of this
work is the use of the most precise and reliable output for the
$\pi\pi$ and $\pi K$ scattering processes, which is a key factor to
attaining high levels of precision and to fix the \rcht\ parameters
given in Eq.~(\ref{eq:chiRTpar}). In the $\pi\pi$ case, we use the recent data
analysis given in~\cite{GarciaMartin:2011cn}.
This analysis incorporates the latest data on $K_{l4}$ decays from
NA48/2~{\cite{Batley:2010zza} as well as constraints from Roy equations
and one-subtracted coupled dispersion relations - or GKPY
(Garcia-Martin, Kaminski, Pelaez and Yndurain) equations.
For the $\pi K$ case, we use the last update of
the Roy-Steiner solutions in~\cite{Buettiker:2003pp}, which
includes input from the $\pi K$ phase-shifts around
1.1 GeV and information on the vector $\pi K$ form-factor from tau
decays.  However, we do not fit to the  $\delta^{\frac{3}{2}1 }$ phase-shift,
as this channel was not considered in the solution of
the Roy-Steiner equations, and it came as prediction of the
scheme. Thus, the subtraction constant $ C_{\frac{3}{2}1}^{K\pi}$ in
Eq.~(\ref{eq:C's}) cannot be determined.

In total, the data set which we are fitting is a compilation of
14 independent channels, shown in Table~\ref{tab:data}, from
the above two independent sources. Additionally, despite using an elaborated
theoretical model to describe these channels, we know
that it contains systematic uncertainties (partial resummation)
or neglected physics (isospin breaking).
A key issue in our fit is therefore how to combine the different
experimental and theoretical inputs in a consistent picture.

The first important point is the choice of the data errors for the
individual channels. Unfortunately, for the $K\pi$ scattering,
the output of the analysis in Ref.~\cite{Buettiker:2003pp} does not provide
any errors. Nevertheless, the input contains some experimental uncertainties,
typically of the 10\% order, and we assume this to be the error for the
$K\pi$-scattering data. The only exception is the $|T^{00}|$ case in
$\pi\pi\to\bar KK$, for which due to its small value above 1 GeV, we also
add 0.1 to the errors. In the case of the $\pi\pi$-scattering channels,
we take the errors from the work of Ref.~\cite{GarciaMartin:2011cn} with
two exceptions: the $\delta^{00}$ phase shift and the $\eta^{00}$ inelasticity.
They are reported to be $1\sim5$\% and $\sim20$\%, respectively. Using these
errors in a combined fit represents certain difficulties as they are
very different compared to all the other channels or the assumptions that
entered the model. On the one hand, the sharp error for the $\delta^{00}$ would
drive the fit to precisely describe this channel on the expense of all the other ones,
especially the $\eta^{00}$. On the other hand, we also do not expect our model
to be accurate to a $<5$\% level. Therefore, to have a more homogeneous
error definition across all channels, we enlarge the reported $\delta^{00}$
error by a factor of $2$ and divide the reported $\eta^{00}$ errors by a
factor of $2\sim2.25$. The enlargement of the $\delta^{00}$ errors is thereby
to be interpreted as a quantitative input of the model uncertainties to the fit.
Concerning the reweighting of $\eta^{00}$, we hope that future,
more precise data, will make this reweighting unnecessary.
In addition, we will see later that the main results of this work are not
significantly affected by this choice. Furthermore, our model does not include isospin
breaking effects, which are known to play a crucial role in the $\delta^{00}$
channel of $\pi\pi\to\pi\pi$ around the region $990$ MeV $<\sqrt{s}$ $<1010$ MeV.
We therefore exclude these data points from the fit.

The second important issue is connected to the used pseudo data points
as their number in each channel is arbitrary. They are analytically generated
from the theoretical analyses carried out in Refs.~\cite{GarciaMartin:2011cn,Buettiker:2003pp},
usually in intervals of $5$ MeV.
To reduce the dependence on this, we normalize each contribution from a given
channel by the number of data points in that channel. The exact $\chi^2$ definition
is given in the next section. In using this normalized approach, the reduction
of the $\eta^{00}$ errors by $2\sim2.25$ is equivalent to give an extra weight of
$4\sim5$ to this channel in the overall $\chi^2$.

With the above settings we will be able to obtain a consistent fit
that homogeneously describes all the 14 channels as well as is compatible
with the theoretical assumptions entering the model.

\begin{table}[tbh]
  \begin{tabular}{cccccc}
    \hline
    \hline
    dataset & range (GeV) & errors & & & $\chi^{\prime 2}$\\ \hline
    $\pi\pi\to\pi\pi$\, \cite{GarciaMartin:2011cn} & && & & \\
    $\delta^{00}$ & [0.28,1.2] & (*) && & 1.6\\
    $\delta^{11}$ & [0.28,1.2] &      && & 1.0\\
    $\delta^{20}$ & [0.28,1.2] &  & & &0.7\\
    $\eta^{00}$ & [0.28,1.2]   & (**) && & 0.5\\
    $\eta^{11}$ & [0.28,1.2]   &  && & 0.0\\
    $\pi\pi\to K\bar K$\, \cite{Buettiker:2003pp} & & & & &\\
    $|T^{00}|$    & [0.99,1.2] & 10\% & & &0.8\\
    $\delta^{00}$ & [0.99,1.2] & $10\%$ & & &0.2\\
    $|T^{11}|$    & [0.99,1.2] & 10\% & & &0.1\\
    $\delta^{11}$ & [0.99,1.2] & 10\% & & &0.0\\
    $\pi K \to \pi K$\, \cite{Buettiker:2003pp} & & && & \\
    $\delta^{\frac12 0}$ & [0.64,1.2] & $10\%$ && & 1.2\\
    $\delta^{\frac12 1}$ & [0.64,1.2] & $10\%$ && & 0.8\\
    $\delta^{\frac32 0}$ & [0.64,1.2] & $10\%$ && & 0.4\\
    $\eta^{\frac12 0}$ & [0.64,1.2] & 0.05 && & 0.0\\
    $\eta^{\frac12 1}$ & [0.64,1.2] & 0.05 && & 0.0\\
    \hline
    \hline
  \end{tabular}
  \caption{Compilation of data included in the fit.
    In the case of the $\pi\pi\to\pi\pi$ channels we take the errors
    as obtained from the CFD parametrization derived in
    \cite{GarciaMartin:2011cn}, with two exceptions: in (*) the errors
    have been multiply by a factor 2 and the region between
    $990\, {\rm MeV}\,<\sqrt{s}<1010$ MeV has been excluded from the
    fit, in (**) the errors have been reduced by a factor of
    $2.25$ (see text for details). The errors on the $\pi\pi\to
    K\bar K$ and $\pi K \to \pi K$ data-sets have been obtained
    from the experimental input uncertainties in
    \cite{Buettiker:2003pp} (see Figs. 2, 4, 8 and 9 of this
    reference), however, in the case of $|T^{00}|$ we also added an
    absolute 0.1 error above 1 GeV. In the last column, we show the contribution
    $\chi^2_{\rm chan}$ of each channel as defined in
    Eq.~(\ref{eq:chi-naive}). \label{tab:data} }
\end{table}

\subsection{The usual $\chi^2$ approach}

On the one hand, a fit to the scattering data of
the $N_{dc} = 14$ channels involves the standard $\chi^2$, defined as
\begin{eqnarray}
  \chi^2 _{\rm exp}& =& N_d \sum_{\rm chan=1}^{N_{dc}} \chi^2_{\rm chan} \;\;\;,\\
  \chi^2_{\rm chan} & =& \frac{1}{n}\sum_{i=1}^{n}\; \left(\frac{O_i^{\rm th}- O_i^{\rm
      exp}}{\Delta O_i^{\rm exp}} \right)^2\, , \label{eq:chi-naive}
\end{eqnarray}
$N_d$ the overall total number of data points. The $(O_i^{\rm
  exp},\Delta O_i^{\rm exp})$ are the fitted observables with their
corresponding errors, and $O_i^{\rm th} = F_i (p, C)$ are our
theoretical descriptions, with $p_j=G_V, c_d , \dots $ parameters
subjected to theoretical conditions and $C_j= C_{00}^ \pi,\dots $
parameters which {\it can only} be determined from data. On the other
hand, we typically expect our calculation to be accurate up to $1/N_c$
corrections. Thus, if a fit turns out to provide numbers completely
different from the large-$N_c$ estimates of Eqs.~(\ref{eq:chiRT1}) and
(\ref{eq:chiRT2}), we will suspect the fit. The optimal situation
would be when the data would have an accuracy able to pin down
reliably all the parameters. Unfortunately, this is {\it not } the
case. Attempts to determine the $C_{IJ}$ subtraction constants in
Eq.~(\ref{eq:C's}) {\it and} the \rcht\ parameters in
Eq.~(\ref{eq:chiRTpar}) (in all 23 parameters) produce multiple minima
with at times quite unreasonable values for the \rcht\ parameters.  In
this case, we are inclined to reject the fit.  If, on the other hand,
the large-$N_c$ constraints are fully implemented\footnote{As done in
  the previous work~\cite{Nieves:2011gb} with a lower energy cut-off
  $\sqrt{s} \lesssim 700 {\rm MeV}$)} and larger error bands are
assumed, the resulting fits are likewise not satisfactory.

We want to obtain a {\it reasonable} fit with {\it sensible}
parameters. Therefore, rather than expecting the fit to
tell us a posteriori whether the parameters are reasonable, we provide a priori a
reasonable guess for the parameters and search for the minimum within
the expected departure of this assumption. In the next section, we
explain how we include the a priori input in the fit.

The existence of multiple solutions is a consequence of several shallow directions
in the parameter space, which it makes impossible to identify a unique
global minimum.
This situation could be solved by including further experimental or theoretical
information. For example, data of those channels that, considered in the coupled channel
formalism, have not been fitted. Unfortunately, this is not the case, and a
Bayesian fit with large-$N_c$ priors becomes a natural and simple way to circumvent this problem.

Guo and Oller~\cite{Guo:2011pa} handle the problem of proliferation
of parameters in a different manner. They took free values for all the resonance
parameters ($p_j$ in our notation) but they invoked to some {\it
  unclear} SU(3) relations among the subtraction constants ($C_j$ in
our notation), and kept just four independent.  However, this is not a
consistent approach. First, the subtraction
constants are not in principle related by SU(3) and should
all be taken as independent (see discussion in \cite{Nieves:1999bx}). Second,
the unconstrained fit of the resonance parameters to data led in some
cases to values in clear contradiction to the large-$N_c$ expectations.\footnote{For instance, a best fit value of 15 MeV for
  $c_d$ was obtained in \cite{Guo:2011pa}, which is around a
factor of three smaller than that of $f_\pi/2$ given in
Eq.~(\ref{eq:chiRT1})} Conversely, the Bayesian approach to be
discussed below, where large $N_c$ constraints are imposed as a probabilistic
prior, does not support the assumptions of Ref. \cite{Guo:2011pa}.

\subsection{The augmented $\chi^2$ approach}

In the Bayesian interpretation, the fitting parameters are actually
random variables which are determined from the existing given data and
a prior probability of finding the parameters, regardless of the
actual measurements under analysis. We shall not dwell into the
philosophical intricacies and use the augmented $\chi^2$ method to fix
the prior
distribution~\cite{Lepage:2001ym,Morningstar:2001je,Schindler:2008fh}.
This approach has successfully been used in lattice QCD to analyze a
number of data with a similar number of parameters. In our case the
situation is slightly different, but we expect the large-$N_c$ limit
to set reasonable ranges on the fitting parameters.

We consider first the theoretical $\chi^2$ defined in Eq.~(\ref{eq:chi2-TH}).
This figure of merit is coherent with the assumption that the prior
probability for the \rcht\ parameters is given by their large-$N_c$
estimate, within a relative $1/N_c$ accuracy (and {\it not} as an uniform
distribution). We use the results of
Eq.~(\ref{eq:chiRT1}) for $G_V,\tilde c_m, c_d,\tilde c_d, d_m , \tilde d_m$ and
Eq.~(\ref{eq:chiRT2}) for $m_{S_0},m_{S_8},m_V$. Thus, we take the
following Gaussian variables normalized to the {\it same} $\Delta p_i=f_\pi/N_c$:
\begin{eqnarray}
\xi_{G_V} &\!\! =&\!\! \frac{\sqrt{2} G_V- f_\pi}{f_\pi/N_c} \, , \qquad \xi_{c_d}\, =\, \frac{2 c_d- f_\pi}{f_\pi/N_c} \, , \qquad \dots
\nonumber \\
\xi_{m_V} &\!\! =&\!\! \frac{ \sqrt{N_c/24\pi^ 2}m_V- f_\pi}{f_\pi/N_c} \, , \qquad\dots
\end{eqnarray}

We provide in addition an {\it a priori} splitting $m_{S_0}-m_{S_8}$ for the scalar
octet and singlet masses, which are equal at large $N_c$:
\begin{equation}
  \xi_{m_{S_0}-m_{S_8}} \; =\;  \sqrt{\frac{N_c}{24\pi^ 2}}\; \frac{m_{S_0}- m_{S_8}}{f_\pi/N_c} \, ,
\end{equation}
and we finally also consider the robust constraint
in the large $N_c$ limit $4c_dc_m\sim
f_\pi^2$, obtained by requiring the $K\pi$ scalar form factor to
vanish in the $t\to \infty$ limit~\cite{Jamin:2000wn},
\begin{equation}
  \xi_{c_dc_m} = \frac{\sqrt{4 c_dc_m} - f_\pi}{f_\pi/N_c}\,.
\end{equation}
Thus, we take into account a total of 10 contributions to construct
$\chi^2_{\rm th}$,
\begin{eqnarray}
  \chi^2_{\rm th} & = & \left[ \xi_{G_V}^2 + \xi_{c_d}^2
    +\xi^2_{c_dc_m} +\xi_{d_m}^2 + \xi^2_{\tilde{c}_d} + \xi^2_{\tilde{c}_m}+\xi_{\tilde d_m}^2\right] \nonumber \\
  &+&\left[ \xi_{m_V}^2 + \xi_{m_{S_8}}^2 + \xi_{m_{S_0}-m_{S_8}}^2\right].
\end{eqnarray}
In addition, we take for the singlet and octet pseudoscalar resonance
masses $m_{P_i}=\sqrt{2}m_{S_i},\,i=0,8$,
see section \ref{sec:sdc}.

The key question now is how to combine $\chi^2_{\rm exp}$ and
$\chi^2_{\rm th}$. Obviously, since we have a small number of
constraints $N_p$ as compared to the number of data or pseudo-data
$N_d$, a direct addition of $\chi^2_{\rm exp}$ and $\chi^2_{\rm th}$
would make the constraints irrelevant. Therefore we will construct a
{\it reduced} $\chi^2$, $\bar \chi^2 \equiv \chi^2 /N $, with a $50\%$
weighting on the data/pseudo-data and the theoretical constraints.

Thus, we  define
\begin{eqnarray}
  \chi^2 _{\rm total}\; =\;  \frac{N_d+N_p}2\; \left(  \frac{\chi^2_{\rm exp}}{N_{d} N_{dc}} +
  \frac{\chi^2_{\rm th}}{N_p} \right)\, .
  \label{eq:chi2tot}
\end{eqnarray}
The additional terms in the total $\chi^2$ impose a penalty for fits
which deviate from the large-$N_c$ expectations by more than
$1/N_c$. This is just a condition on the naturalness of parameters,
based on a simple large-$N_c$ estimate. Of course, the values we are
taking as a reference are based just on the single-resonance
approximation, and this is precisely why one should not attach
exaggerated significance to the detailed accuracy of the reasonable
fit. The opposite situation, the impossibility of performing a
successful fit would signal a serious drawback of the whole framework,
including the usefulness of the short-distance constraints in
meson-meson scattering.

Furthermore, we have checked our approach against the weighting choice
of Eq.~\eqref{eq:chi2tot}. We found that the parameters in Table~\ref{tab:fit-results} do not depend
strongly on this particular setting as long as the fit starts
in the vicinity of the respective minimum. That is, the augmentation
of Eq.~\eqref{eq:chi2tot} is needed to isolate the physical minimum of Table~\ref{tab:fit-results}
from all the unphysical local minima, but after having found it
the corresponding parameters do not strongly depend on the specific choice of Eq.~\eqref{eq:chi2tot}.

\subsection{Results}

\begin{table}[h]
\begin{tabular}{cccccc}
\hline
\hline
Parameter & Large $N_{c}$ & Fit  &  &  & Fit \tabularnewline
$G_{V}$ & $65.3$ & $63.2\,(1)$ &  & $C_{00}^{\pi\pi}$ & $-0.0209\,(2)$\tabularnewline
$c_{d}$ & $46.2$ & $39.8\,(1)$ &  & $C_{00}^{K\bar K}$ & $-0.0085\,(3)$\tabularnewline
$\tilde c_{d}$ & $26.7$ & $20.7\,(3)$ &  & $C_{00}^{\eta\eta}$ & $0.0060\,(2)$\tabularnewline
$c_{m}$ & $46.2$ & $41.1\,(1)$ &  & $C_{11}^{\pi\pi}$ & $-0.0279\,(6)$\tabularnewline
$\tilde c_{m}$ & $26.7$ & $18.9\,(9)$ &  & $C_{11}^{K\bar K}$ & $-0.0100\,(5)$\tabularnewline
$d_{m}$ & $32.7$ & $29.8\,(1)$ &  & $C_{20}^{\pi\pi}$ & $-0.0561\,(5)$\tabularnewline
$\tilde d_{m}$ & $18.9$ & $19.3\,(7)$ &  & $C_{\frac{1}{2}0}^{K\pi}$ & $-0.0029\,(3)$\tabularnewline
$m_{V}$ & $821$ & $805\,(2)$ &  & $C_{\frac{1}{2}0}^{K\eta}$ & $-0.0165\,(14)$\tabularnewline
$m_{S_{0}}$ & $821.0$ & $808.9\,(4)$ &  & $C_{\frac{1}{2}1}^{K\pi}$ & $0.0380\,(20)$\tabularnewline
$m_{S_{8}}$ & $821$ & $1279\,(9)$ &  & $C_{\frac{1}{2}1}^{K\eta}$ & $-0.0240\,(20)$\tabularnewline
$\mu_{V}$ &  & $645\,(4)$ &  & $C_{\frac{3}{2}0}^{K\pi}$ & $-0.0394\,(14)$\tabularnewline
$\mu_{S}$  &  & $274\,(10)$ &  & $C_{\frac{3}{2}1}^{K\pi}$ &$-$ \\
\hline
\hline
\end{tabular}
\caption{Parameters (in MeV) of the \rcht\ Lagrangian determined
  from phenomenology~\cite{Ecker:1988te,Ecker:1989yg} or
  short-distance constraints~\cite{Pich:2002xy} and the resulting
  values from the combined fit. Resonance masses and
  saturation scales are also given in MeV, while subtraction constants
are dimensionless.}\label{tab:fit-results}
\end{table}
\begin{table*}[t!]
  \begin{tabular}{cccccccc}
    \hline
    \hline
    $I$&$J$ & & $\sqrt{s_R}$ [MeV]& $|g_{\pi\pi,\pi K}|$ [GeV] &$|g_{\bar KK}|$ [GeV] & PDG
    ~\cite{Beringer:1900zz} $\sqrt{s_R}$ [MeV]
    & GKPY/RS~\cite{GarciaMartin:2011jx,DescotesGenon:2006uk} $\sqrt{s_R}$ [MeV]\\\hline
    0&0& $f_0(500)$ or $\sigma$& $458(2) -i\,264(3)$ &$3.3(1)$ &$0.9(1)$ &
    $400\sim 550 -i\,[200\sim 350]$ &  $457^{+14}_{-13} -i\, 279^{+11}_{-7}$ \\
    0&0& $f_0(980)$   & $1001(4) -i\,24(2)$ & $2.2(1)$ &$4.2(1)$ &
    $990\pm 20 -i\,[20 \sim 50]$ & $996(7) -i\,25^{+10}_{-6}$ \\
    0&0 & & $1211(64) -i\,332(78)$ &$3.3(5)$ &$2.6(2)$ &&\\
    0&0 & & $1449(29) -i\,61(12)$ &$3.1(2)$ &$0.3(1)$ &&\\
    1&1 & $\rho(770)$ &$761(3)-i\,73(2)$ &$6.0(1)$ &$3.9(1)$ & $772.3(9)-i\,75.1(8)^*$&$763.7^{+1.7}_{-1.5} -i\,73.2^{+1.0}_{-1.1}$\\
    $1/2$&0& $K_0^*(800)$  or $\kappa$& $684(4) -i\,260(4)$ &$4.2(1)$ & & $682(29) -i\,273(12)$ & $659(13) -i\,278(12)$\\
    $1/2$&1 & $K^*(892)$ &$897(5) -i\,24(2)$&$5.4(3)$& &$896.1(19) -i\,23.7(3)^*$\\
    \hline
    \hline
  \end{tabular}
\caption{Pole positions and resonance couplings found in this work with the
  best fit parameters compiled in Table~\ref{tab:fit-results}.
  All statistical uncertainties are defined on a 68\% confidence-level.
  For the resonances marked with $(^*)$, the PDG
  quote Breit-Wigner (BW) parameters $s_{\rm pole}=M_{BW}^2-
  iM_{BW}\Gamma_{BW}$, from which we have computed the corresponding
  pole positions. For the $\rho(770)$ case we show the ``Neutral only,
  other reactions'' average values, whereas for the $K^*(892)$ we give
  the ``Neutral only'' average mass and width. Note that the
  dispersive data analysis of Ref.~\cite{Caprini:2005zr}, based in the
  Roy equations, predict $\sqrt{s_\sigma}= (441^{+16}_{-8} -i\,
  272^{+9}_{-12})$ MeV. In addition, the  $|g_{\pi\pi}|$ GKPY dispersive determination
  given in \cite{GarciaMartin:2011jx} for the $f_0(500)$, $f_0(980)$ and $\rho(770)$
  are $3.59^{+0.11}_{-0.13}$, $2.3\pm 0.2$ and $6.01^{+0.04}_{-0.07}$ respectively.}\label{tab:fit-poles}
\end{table*}
\begin{figure*}
\begin{center}
\includegraphics[scale=0.5]{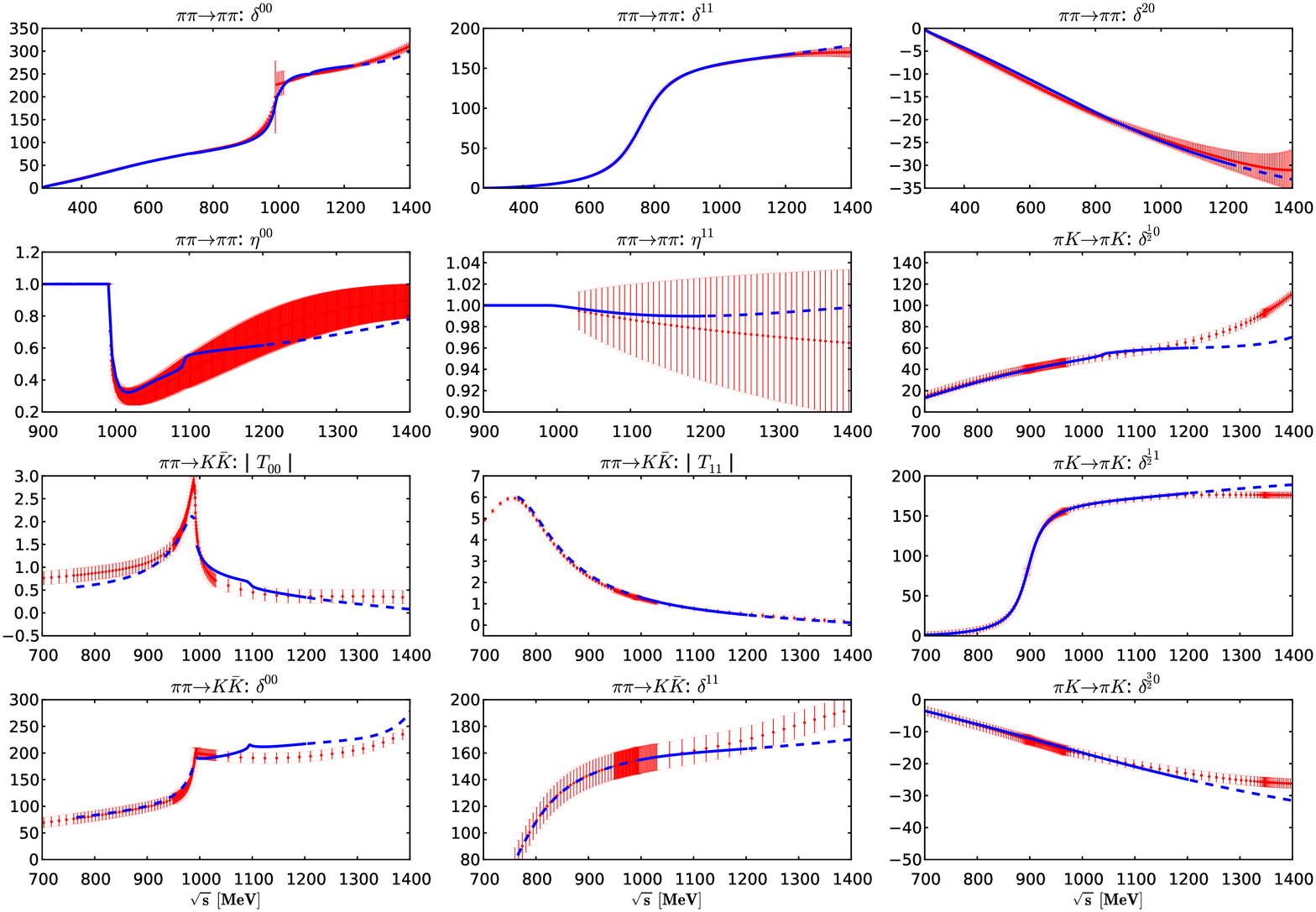}
\end{center}
\caption{\label{fig:fits} Results for meson-meson scattering
  observables as a function of the CM energy $\sqrt{s}$ (in MeV).  We
  display the pseudo-data, with their assumed uncertainties, included in the
  fit (see Table~\ref{tab:data}).
  Solid lines represent fitted curves,  dashed (blue) lines represent
  non-fitted curves.}
\end{figure*}
The values of the fitted parameters are presented in
Table~\ref{tab:fit-results} and the results for scattering properties
are depicted in Fig.~\ref{fig:fits}, where solid (blue) lines
represent fitted curves. The results for non-fitted data are depicted
as dashed (blue) lines. In the Bayesian approach, errors on the
parameters are estimated as {\it mean} values, i.e., integrating the
likelihood with respect to the fitting parameters, but when the total
$\chi^2$ is large (not the $\chi^2/{\rm d.o.f.})$, as it is the case
here, a saddle point approximation can be used. This is just
equivalent to determine them by the standard covariance matrix
inversion method applied in our case to Eq.~(\ref{eq:chi2tot}).

As expected, the fit below $\Lambda_R = 1.2$~GeV is successful with
reasonable resonance parameters motivated by large-$N_c$ constraints
(when states with mass above $\Lambda_R$ are disregarded). Indeed, as
it can be seen in the Table~\ref{tab:fit-results}, all \rcht\, parameters
turn out to be in agreement, within the typical 30\% uncertainty,
with the large-$N_c$ expectations.
The only exception is the value of $m_{S_8}\sim 1280$ MeV, which
lies outside of the $N_c$ expected region.
However, the mean between the two scalar masses, $m_{S_8}$ and $m_{S_0}$,
satisfies this constraint.

The achieved description  for all considered 14 {\it pseudo-data}
channels is quite good, as can be appreciated in
Fig.~\ref{fig:fits}. This is even more relevant, taking into account
that the comparison is being made with the quite precise
output obtained from the data constrained Roy-GKPY and Roy-Steiner
analyses carried out in
Refs.~\cite{GarciaMartin:2011cn,Buettiker:2003pp}, which provide the
most reliable information currently available in
the literature on the various scattering amplitudes.

Next, we discuss the poles found in the SRS of the amplitudes. The SRS
of the $T$ matrix is determined by the definition of the loop function
${\mathcal{I}}_{0}^{\phi\phi'}(s)$. As mentioned above, we
use the Eq.~(A13) of Ref.~\cite{Nieves:2001wt}.  Masses and widths of
the dynamically generated resonances are determined from the positions
of the poles, $s_R$, in the SRS of the corresponding scattering
amplitudes in the complex $s$ plane.  Since in the SRA amplitudes we have
explicitly incorporated one vector and two scalar poles, we expect at
least these poles to appear in the appropriate sectors. However,
because of the re-summation in Eq.~(\ref{eq:tunita}), the pole
positions will change with respect to those of
the bare ones ($s=m^2_V, m^2_{S_0}$ and $m^2_{S_8}$) and the resonances will acquire a
width that accounts for their two-meson decay.
This change is specially relevant for the $f_0(500)$ or $\sigma$ meson,
where pion loops dominate at $N_c=3$, producing a state very deep on the
complex plane, with a mass around 450 MeV and a width around 520 MeV.
As studied in~\cite{Cohen:2014vta}, this effect is due to a strong
cancellation between different $N_c$ orders, which makes it difficult to analyze
its properties just from the pure $1/N_c$ expansion.
In fact, there are several works in the literature~\cite{Nieves:2011gb,
Pelaez:2003dy,RuizdeElvira:2010cs,Guo:2011pa,Guo:2012yt,Guo:2012ym},
in which the $f_0(500)$ fades away on the complex plane as $N_c$ increases,
being made just of meson-meson loops,
and having then, no relation with the scalar poles included in the Lagrangian.

In addition, as we will see, some other poles are generated as well in the SRS of the
scattering amplitudes. The results are presented in the
Table~\ref{tab:fit-poles}. We find a quite good description of the
$f_0(500),f_0(980),K^*_0(800), \rho(770)$ and $K^*(892)$ resonances,
with masses and widths that compare rather well with the averaged ones
compiled in the Review of
Particle Properties~\cite{Beringer:1900zz}.  Furthermore, since our
results have been obtained by fitting the pseudo-data values obtained
in the $\pi\pi$~\cite{GarciaMartin:2011cn} and $\pi
K$~\cite{Buettiker:2003pp} dispersive analyses, we also include in the
last row of Table~\ref{tab:fit-poles}, the dispersive determinations
obtained from these schemes~\cite{GarciaMartin:2011jx,DescotesGenon:2006uk}.
The agreement is also remarkable. We want to note that the properties of all
these dynamically generated states are not significantly affected by the
employed re-weighting of the $\delta^{00}$ and $\eta^{00}$ channels.

Let us now pay  attention  to the complex pole structure
of the scalar-isoscalar sector, depicted in Fig.~\ref{fig:poles}. The
$f_0(500)$ and $f_0(980)$ resonances are clearly visible. Besides,
there appear two additional poles, which are placed above  $\Lambda_R
= 1.2$~GeV. Perhaps, the lower one could have some relation with the
$f_0(1370)$ state. This will agree with the findings of
Ref.~\cite{Albaladejo:2008qa}, where the $f_0(1370)$ is identified as a pure octet state
not mixed with the glueball. The chiral unitary approaches,
supplemented by the inclusion of vector mesons,  of
Refs.~\cite{Geng:2008gx, GarciaRecio:2010ki,  Garcia-Recio:2013uva}
seem to give also support to this hypothesis. Nevertheless, the exact
position of the higher poles depends much more on the choice of the
merit function which is being minimized. Especially, the position
of the $f_0(1370)$ depends strongly on how the comparatively imprecise $\eta^{00}$
pseudo data is included in the fit.
In addition, these states are located above the scale of the first resonance
multiplet considered, and then, are strongly dependent on those heavier
states integrated out in the starting Lagrangian.
Therefore, these heavier poles cannot be properly described within the framework
used in this work and are included only for completeness.

Finally in Table~\ref{tab:fit-poles}, we also  provide
for each resonance its coupling to the fitted channels,
($\pi K$ in the case of the $K^*_0(800)$ and $K^*(892)$,
and $\pi\pi$ and $KK$ for the others),
defined from its pole residue as,
\begin{equation}
  g_Ag_B=\lim_{s\rightarrow s_{\rm pole}}(s-s_{\rm pole})T_{IJ}^{A\to B}(s)(2J+1)/(2p)^{2J},
\end {equation}
where $p$ is the center-of-mass-system momentum of the corresponding process.

\begin{figure}
  \begin{center}
    \includegraphics[scale=0.35]{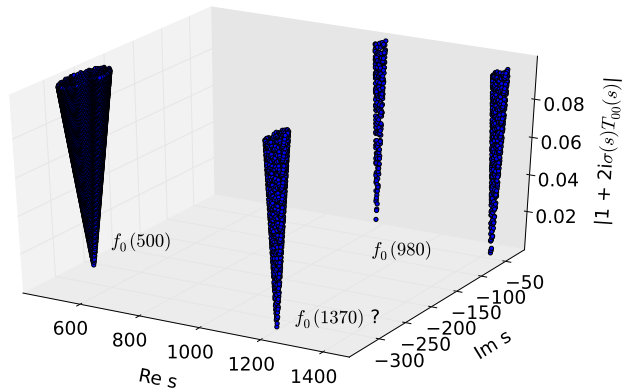}
  \end{center}
  \caption{\label{fig:poles} Poles found in the SRS of scalar-isoscalar
    elastic $\pi\pi$ amplitude. The $f_0(500)$ and $f_0(980)$ are clearly
    visible. }
\end{figure}

\section{Conclusions}
\label{sec:concl}

Within a unitarized coupled-channel approach, we have analyzed the
scattering of the pseudo-scalar mesons
for $\sqrt{s} \le 1.2$~GeV, where all
two-body scattering channels are open. Our amplitudes contain one-loop
(NLO) \chpt\ and tree-level (LO) large-$N_c$ pieces, with 24 fitting
parameters.  Given the lack of very precise experimental  data, in this
work we have used the most precise and reliable output for the
$\pi\pi$ and $\pi K$ scattering processes from the Roy-GKPY and
Roy-Steiner analyses carried out in
Refs.~\cite{GarciaMartin:2011cn,Buettiker:2003pp}. This is a major
novelty of this work, since for the very first time, these two sets of
data or pseudo-data have been simultaneously used to constrain
the LECs. Indeed, it has been a key factor to
attaining high levels of precision and to fix the
\rcht\ parameters. However, not all of the pseudo-data have
experimentally inherited uncertainties, and hence an educated guess
in defining the used merit function has been made.

While our model contains important features of the true solution and
we optimized it by minimizing the discrepancies with {\it experiment},
the large number of best fit parameters made the task of performing
the fit difficult. We faced a quite complex structure
of the parameters manifold, which has many resemblances to a
multidimensional egg-box. The proliferation
of undetermined LECs made direct fits rather elusive and quite often we were
driven to unreasonable parameter values, which suggested
rejecting the fit. Under these circumstances, we have adopted the Bayesian
point of view of making the natural assumption that the \rcht\ parameters
take their large-$N_c$ estimated values within an expected $1/N_c$ uncertainty.

The main outcome of the present study is that a rather good
description of the data can be achieved with natural values of the
parameters and considering the nominal expected accuracy of the
calculations. This is a non-trivial result, and an important
ingredient for this success is the allowance for systematic deviations
in all parameters where the large-$N_c$ expansion is expected to
provide corrections of ${\cal O}(1/N_c)$. The predictions
compiled in Table~\ref{tab:fit-poles} for pole positions and couplings of
the lowest-lying dynamically generated resonances, which show a  nice
agreement with the most precise current determinations, are an example of this
success.

\section*{Acknowledgments}

We thank B. Moussallam for the results of the update of his
Roy-Steiner dispersive analysis of $\pi K$ scattering and
Z.H. Guo for discussions on the approach
employed in Ref.~\cite{Guo:2011pa}. We also acknowledge useful discussions with M. Albaladejo.
This work has been supported in part by the Spanish Government and
ERDF funds from the EU Commission [grants FIS2011-24149,
  FIS2011-28853-C02-01, FIS2011-28853-C02-02, FPA2011-23778 and
  CSD2007-00042 (Consolider Project CPAN)], Generalitat Valenciana
[grants PROMETEO/2009/0090 and PROMETEOII/2013/007], Junta de
Andaluc{\'{\i}}a [grant FQM225], the DFG (SFB/TR 16, ``Subnuclear
Structure of Matter'') and the EU Hadron-Physics3 project [grant
  283286].

\appendix
\section{SRA \rcht\, and one-loop \chpt\, amplitudes}\label{sub:APP-Iso-spin-projected-amplitudes}

The ${\cal O}(p^4)$ \chpt\, amplitudes ($T_{IJ}^{\chi PT}(s)$) used
through this work are obtained from
Ref.~\cite{GomezNicola:2001as}. There, and assuming crossing
symmetry, the isospin projected amplitudes for every
possible process involving $\pi,K,\eta$ mesons can be
found. Next, the  individual
contributions with total angular momentum $J$ are calculated using
Eq.~(\ref{eq:IJ-amplitudes}).
Possible $\eta-\eta'$ mixing effects are not taken into
account, and thus the $\eta$ meson is  identified with the isospin
singlet ($\eta_8$) of the octet of Goldstone bosons. In addition, the
normalizations used here are such that our amplitudes differ
in one sign with those given in  \cite{GomezNicola:2001as}.
\begin{itemize}
\item $\pi\pi\to \pi\pi$: There is only
one independent amplitude, $T(s,t,u)$, that is taken to be the
$\pi^+\pi^-\to \pi^0\pi^0$, which at one loop in \chpt\, is given in
Eq.~(B4) of Ref.~\cite{GomezNicola:2001as}. Linear combinations of
$T(s,t,u)$, $T(t,s,u)$ and $T(u,t,s)$ provide the isoscalar, isovector
and isotensor amplitudes (see text above Eq.(12) in
\cite{GomezNicola:2001as}).

\item $K\pi\to K\pi$: Crossing symmetry allows us to write the $I=1/2$
  amplitude (Eq.(12) in \cite{GomezNicola:2001as}) in terms of the $I=3/2$ $K^+\pi^+\to K^+\pi^+$
  one, which is given in Eq.~(B5) of Ref.~\cite{GomezNicola:2001as}.

\item $K\bar K\to K \bar K$: In this case, the two isospin amplitudes
can be expressed in terms of the $\bar K^0 K^0 \to K^+K^-$
amplitude (see Eq.~(25) of Ref.~\cite{Guo:2011pa}), which
  expression to one loop can be obtained  from Eq.~(B8) of
  \cite{GomezNicola:2001as}. Note that this latter equation suffers
  from a typo and there,  it turns out to be the amplitude   $ K^0 \bar K^0 \to
  K^+K^-$ the one which is given instead of the $\bar K^0 K^0 \to K^+K^-$ one.

\item $K\bar K\to \pi \pi$: Thanks to crossing symmetry, the
  amplitudes in this sector are determined by the $I=3/2$ $K^+\pi^+\to
  K^+\pi^+$ one.

\item $K\eta \to K \eta$: This is a pure $I=1/2$ process. The
one-loop amplitude is given in Eq.~(B2) of
Ref.~\cite{GomezNicola:2001as}.

\item  $\bar K K \to \eta \eta$: This is an $I=0$ process that using
crossing symmetry can be obtained from the previous amplitude.

\item $K\eta \to K\pi$: This is also an $I=1/2$ process, and the one
loop expression for $\bar K^0\eta \to K^0\pi^0$ can be found in
Eq.~(B3) of Ref.~\cite{GomezNicola:2001as}.

\item $K K \to \pi\eta$: This $I=1$ process is related to the
$\bar K^0\eta \to K^0\pi^0$ amplitude by crossing symmetry.

\item $\pi\eta \to \pi \eta$: This is a pure $I=1$ isospin process,
  and its amplitude is given in Eq.~(B6) of Ref.~\cite{GomezNicola:2001as}.

\item $\pi\pi\to \eta \eta$: This amplitude is determined from the
  previous one by crossing.

\item $\eta \eta \to \eta\eta$: This pure $I=0$ amplitude
at one loop in \chpt\, is given in Eq.~(B1) of Ref.~\cite{GomezNicola:2001as}.
\end{itemize}

\begin{figure*}
  \begin{centering}
    \includegraphics[scale=0.29]{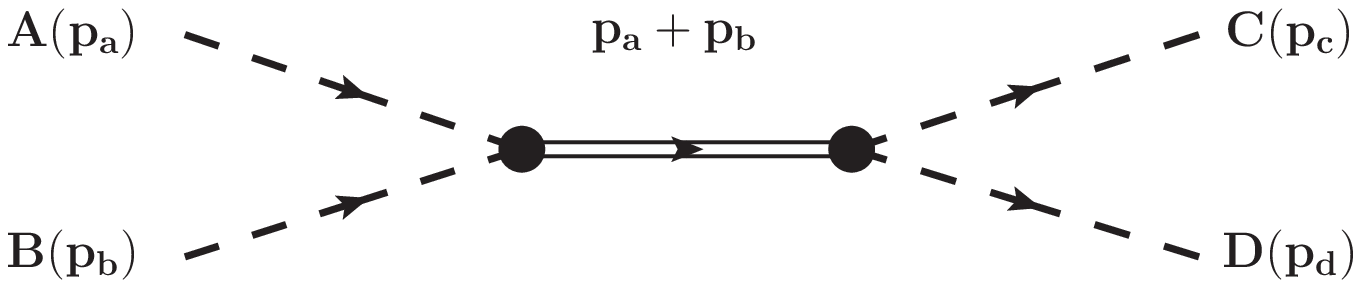}~~~~\includegraphics[scale=0.29]{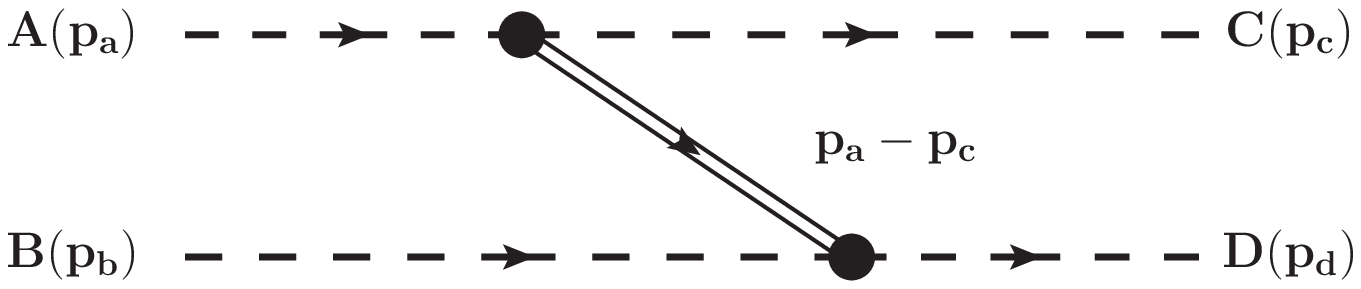}~~~~\includegraphics[scale=0.29]{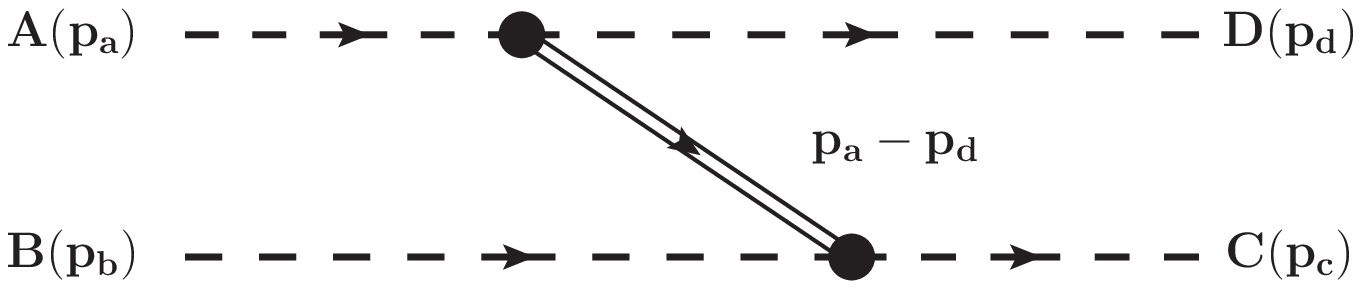}~~~~\includegraphics[scale=0.29]{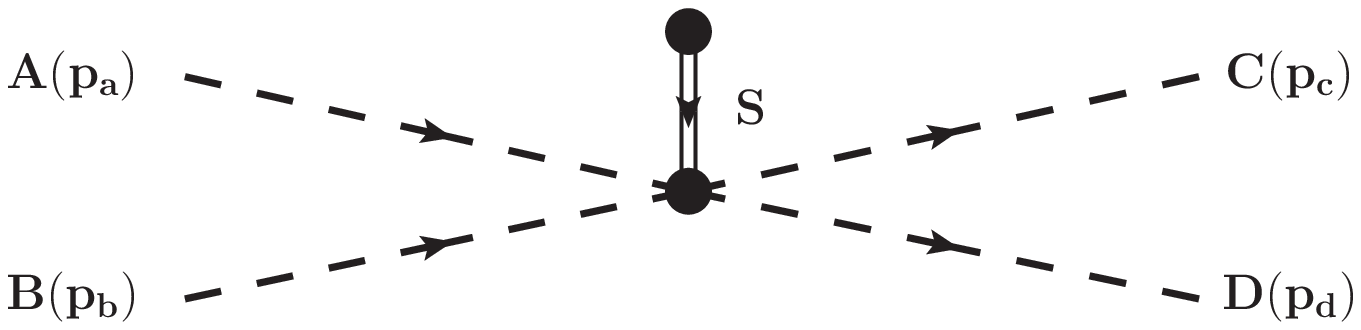}\\
    \hspace{1cm}\\
    \hspace{1cm}\\
    \includegraphics[scale=0.29]{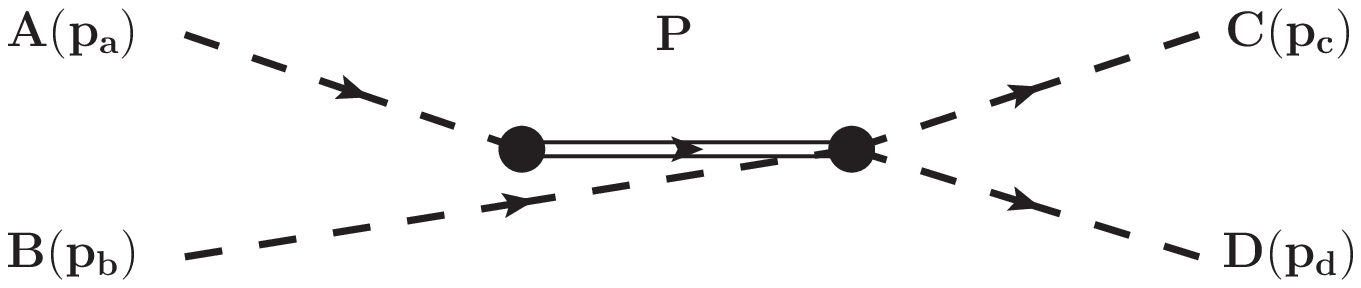}~~~~\includegraphics[scale=0.29]{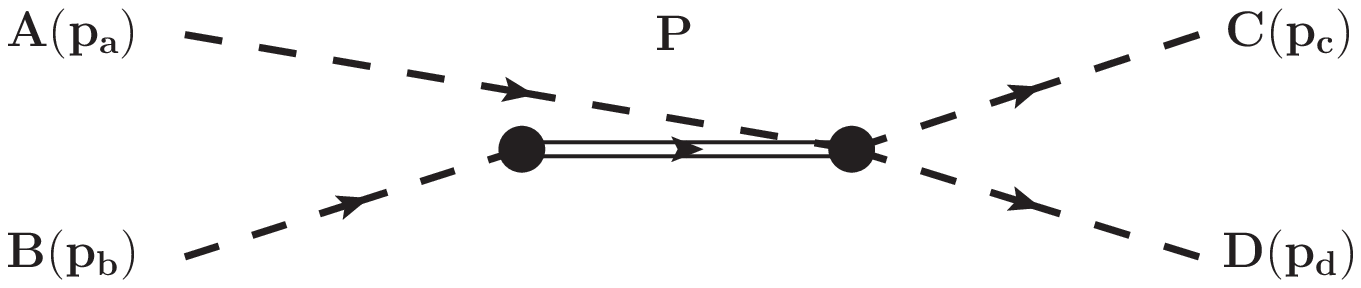}~~~~\includegraphics[scale=0.29]{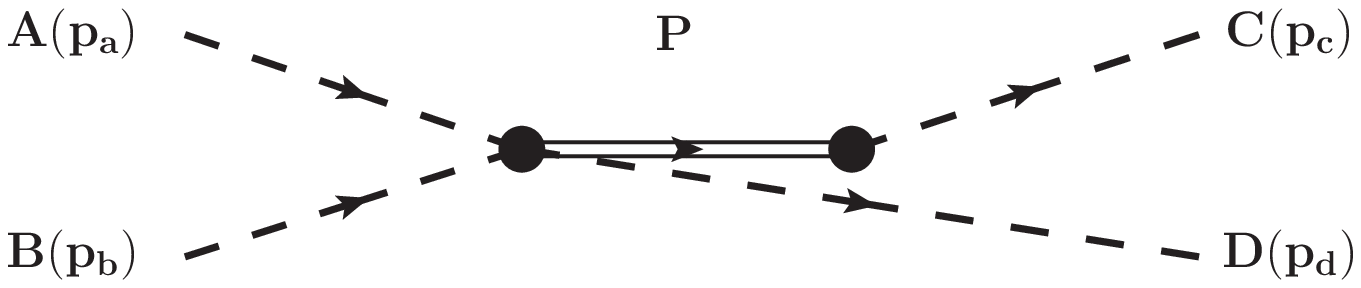}~~~~\includegraphics[scale=0.29]{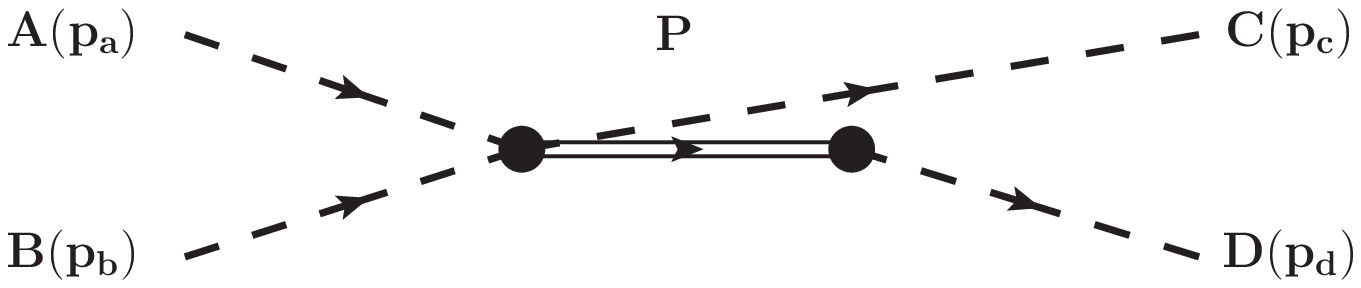}
    \par\end{centering}
    \caption{\label{fig:Meson-meson-scattering-resonance}Resonance
      contributions to the pseudo-scalar
      meson-meson scattering . The dotted lines denote pseudo-scalar mesons,
      double lines the intermediate resonances and the black dot their interaction.
      The first row contains the diagrams with intermediate vector and scalar
      resonances in the $s$, $t$ and $u$ channel graphs. The pseudo-scalar resonances
      contributions are shown in the second row.}
\end{figure*}

Next, we compile the \rcht\, amplitudes $T^{\rm SRA}(s,t,u)$,
within the SRA, for the independent processes mentioned above.
The different Feynman diagrams corresponding to the resonance exchange amplitudes are
illustrated in Fig.~\ref{fig:Meson-meson-scattering-resonance}.
Note that, as it has been anticipated in Eq.~\eqref{eq:unita-sra}, tree level
amplitudes are also included in $T^{\rm SRA}(s,t,u)$.
In addition, resonances also contribute indirectly through the diagrams of
Fig.~\ref{fig:R-contribution2Selgenergy} which modify at $\mathcal{O}\left(p^{4}\right)$
the pion-decay constant and the pseudo-scalar meson self-energy $\Sigma^R_{\phi=\pi,K,\eta}(p^2)$,
and consequently, pseudo-scalar masses and wave-function renormalization constants.
The scalar and pseudo-scalar resonances contribution to the pion-decay constant and self
energy renormalization reads,
\begin{widetext}
  \begin{eqnarray}
    f_{\pi} &=&f\left(1+\frac{8}{3}\frac{c_{m}c_{d}}{f_{0}^{2}m_{S_8}^{2}}\left(m_{\pi}^{2}-m_{K}^{2}\right)+4\frac{\tilde{c}_{m}\tilde{c}_{d}}{f_{0}^{2}m_{S_0}^{2}}\left(2m_{K}^{2}+m_{\pi}^{2}\right)\right),\nonumber\\
    \Sigma_{\pi}^{S}\left(p^{2}\right)&=&\frac{-16c_{m}\left(m_{\pi}^{2}-m_{K}^{2}\right)}{3f_{0}^{2}m_{S_8}^{2}}\left[c_{d}p^{2}-c_{m}m_{\pi}^{2}\right]-\frac{8\tilde{c}_{m}\left(2m_{K}^{2}+m_{\pi}^{2}\right)}{f_{0}^{2}m_{S_0}^{2}}\left[\tilde{c}_{d}p^{2}-\tilde{c}_{m}m_{\pi}^{2}\right],\nonumber\\
    \Sigma_{K}^{S}\left(p^{2}\right) &=& \frac{-8c_{m}\left(m_{\pi}^{2}-m_{K}^{2}\right)}{3f_{0}^{2}m_{S_8}^{2}}\left[-c_{d}p^{2}+c_{m}m_{K}^{2}\right]-\frac{8\tilde{c}_{m}\left(2m_{K}^{2}+m_{\pi}^{2}\right)}{f_{0}^{2}m_{S_0}^{2}}\left[\tilde{c}_{d}p^{2}-\tilde{c}_{m}m_{K}^{2}\right],\\
    \Sigma_{\phi=\pi,K}^{P}\left(p^{2}\right)&=& 8\frac{d_{m}^{2}m_{\phi}^{4}}{f_{0}^{2}}\frac{1}{p^{2}-m_{P_8}^{2}},\nonumber\\
    \Sigma_{\eta}^{S}\left(p^{2}\right)& = & \frac{4}{3}\Sigma_{K}^{S}\left(p^{2}\right)-\frac{1}{3}\Sigma_{\pi}^{S}\left(p^{2}\right).\nonumber
  \end{eqnarray}

  \begin{figure}[h]
    \begin{centering}
      \includegraphics[scale=0.32]{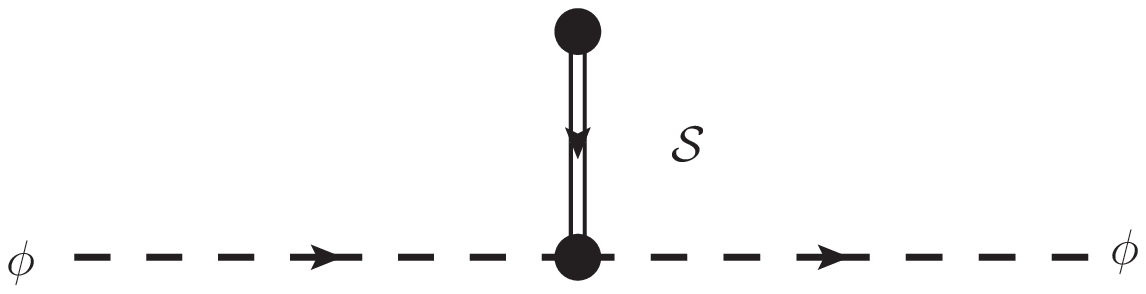}~~~~\includegraphics[scale=0.32]{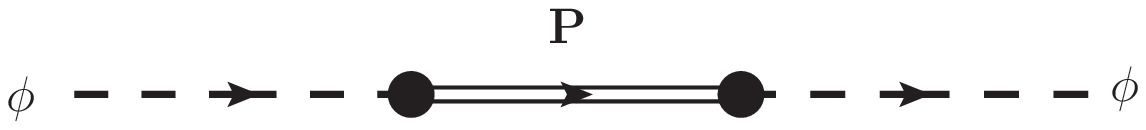}~~~~\includegraphics[scale=0.32]{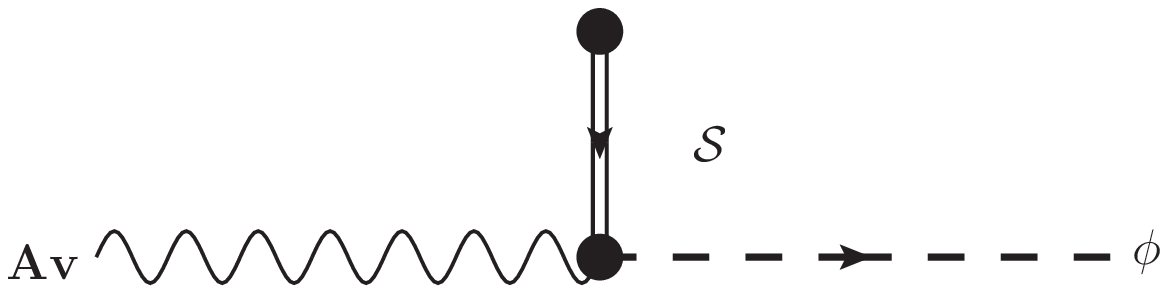}~~~~
      \par\end{centering}
      \caption{\label{fig:R-contribution2Selgenergy} Scalar and pseudo-scalar resonance contributions
        to the pseudo-scalar meson self-energy and scalar one to the pion decay constant $f_{\pi}$.}
  \end{figure}

  Therefore, taking into account all these contributions, the $T^{\rm SRA}(s,t,u)$ amplitude for the channel $\pi^{+}\pi^{-}\to\pi^{0}\pi^{0}$ is given by:
  \begin{eqnarray}
    T^{\rm SRA}(s,t,u)&=&\frac{1}{f_{\pi}^{2}}\left(m_{\pi}^{2}-s\right)+\frac{G_{V}^{2}}{f_{\pi}^{4}}\left(\frac{t\left(s-u\right)}{t-m_{V}^{2}}+\frac{u\left(s-t\right)}{u-m_{V}^{2}}\right)+\frac{2\left(2c_{m}m_{\pi}^{2}+c_{d}\left(-2m_{\pi}^{2}+s\right)\right)^{2}}{3f_{\pi}^{4}\left(s-m_{S_8}^{2}\right)}\nonumber \\
    && +\frac{4\left(2\tilde{c}_{m}m_{\pi}^{2}+\tilde{c}_{d}\left(-2m_{\pi}^{2}+s\right)\right)^{2}}{f_{\pi}^{4}\left(s-m_{S_0}^{2}\right)}-\frac{8m_{\pi}^{4}d_{m}^{2}}{f_{\pi}^{4}\left(m_{\pi}^{2}-m_{P_8}^{2}\right)}.
  \end{eqnarray}
  For the channel $K^{+}\pi^{+}\to K^{+}\pi^{+}$,
  \begin{eqnarray}
    T^{\rm SRA}(s,t,u)&=&\frac{1}{2f_{\pi}^{2}}\left(s-m_{K}^{2}-m_{\pi}^{2}\right)
    -\frac{G_{V}^{2}}{2f_{\pi}^{4}}\left\{\frac{t(s-u)}{t-m_{V}^{2}}+\frac{\left(t-2m_{K}^{2}\right)\left(t-2m_{\pi}^{2}\right)-(m_{K}^{2}+m_{\pi}^{2}-s)^{2}}{u-m_{V}^{2}}\right\}\nonumber \\
    && +\frac{4\left(2(-\tilde{c}_{d}+\tilde{c}_{m})m_{K}^{2}+\tilde{c}_{d}t\right)\left(2(-\tilde{c}_{d}+\tilde{c}_{m})m_{\pi}^{2}+\tilde{c}_{d}t\right)}{f_{\pi}^{4}\left(t-m_{S_0}^{2}\right)}+\frac{\left((c_{d}-c_{m})\left(m_{K}^{2}+m_{\pi}^{2}\right)-c_{d}u\right)^{2}}{f_{\pi}^{4}\left(u-m_{S_8}^{2}\right)}\\
    && -\frac{\left(2(c_{d}-c_{m})m_{K}^{2}-c_{d}t\right)\left(2(c_{d}-c_{m})m_{\pi}^{2}-c_{d}t\right)}{3f_{\pi}^{4}\left(t-m_{S_8}^{2}\right)}
    -\frac{4d_{m}^{2}m_{\pi}^{2}m_{K}^{2}}{f_{\pi}^{4}}\left(\frac{1}{m_{K}^{2}-m_{P_8}^{2}}+\frac{1}{m_{\pi}^{2}-m_{P_8}^{2}}\right)\nonumber\\
    &&+\frac{c_{m}\left(m_{K}^{2}-m_{\pi}^{2}\right)\left(3c_{m}\left(m_{K}^{2}-m_{\pi}^{2}\right)+c_{d}\left(4m_{K}^{2}+4m_{\pi}^{2}-5s+t+u\right)\right)}{3f_{\pi}^{4}m_{S_8}^{2}}.\nonumber
  \end{eqnarray}

  For $K^{0}\overline{K}^{0}\to K^{+}K^{-}$ the SRA amplitude reads,
  \begin{eqnarray}
    T^{\rm SRA}(s,t,u)&=&\frac{u-2m_{K}^{2}}{2f_{\pi}^{2}}+\frac{G_{V}^{2}}{2f_{\pi}^{4}}\left(\frac{s\left(t-u\right)}{s-m_{V}^{2}}+\frac{t\left(s-u\right)}{t-m_{V}^{2}}\right) +\frac{4\left(2\tilde{c}_{m}m_{K}^{2}+\tilde{c}_{d}\left(-2m_{K}^{2}+s\right)\right)^{2}}{f_{\pi}^{4}\left(s-m_{S_0}^{2}\right)}\\
    && -\frac{\left(2c_{m}m_{K}^{2}+c_{d}\left(-2m_{K}^{2}+s\right)\right)^{2}}{3f_{\pi}^{4}\left(s-m_{S_8}^{2}\right)}+\frac{\left(2c_{m}m_{K}^{2}+c_{d}\left(-2m_{K}^{2}+t\right)\right)^{2}}{f_{\pi}^{4}\left(t-m_{S_8}^{2}\right)}\nonumber\\
    &&-\frac{8d_{m}^{2}m_{K}^{4}}{f_{\pi}^{4}\left(m_{K}^{2}-m_{P_8}^{2}\right)}+\frac{4c_{d}c_{m}\left(m_{K}^{2}-m_{\pi}^{2}\right)\left(2m_{K}^{2}-u\right)}{f_{\pi}^{4}m_{S_8}^{2}}\nonumber,
  \end{eqnarray}
  whereas for the process $\overline{K}^{0}\eta\to\overline{K}^{0}\eta$,
  \begin{eqnarray}
    T^{\rm SRA}(s,t,u)&=&\frac{-9t+6m_{\eta}^{2}+2m_{\pi}^{2}}{12f_{\pi}^{2}}-\frac{3G_{V}^{2}}{4f_{\pi}^{4}}\left\{\frac{\left(\left(t-2m_{K}^{2}\right)\left(t-2m_{\eta}^{2}\right)-(m_{K}^{2}+m_{\eta}^{2}-u)^{2}\right)}{\left(s-m_{V}^{2}\right)}\right .\nonumber \\
    && \left.+\frac{\left(\left(t-2m_{K}^{2}\right)\left(t-2m_{\eta}^{2}\right)-(m_{K}^{2}+m_{\eta}^{2}-s)^{2}\right)}{\left(u-m_{V}^{2}\right)}\right\}
    -\frac{\left(c_{m}\left(-5m_{K}^{2}+3m_{\pi}^{2}\right)+c_{d}\left(m_{\eta}^{2}+m_{K}^{2}-s\right)\right)^{2}}{6f_{\pi}^{4}\left(m_{S_8}^{2}-s\right)}\nonumber \\
    && +\frac{\left(2(c_{d}-c_{m})m_{K}^{2}-c_{d}t\right)\left(2c_{m}\left(8m_{K}^{2}-5m_{\pi}^{2}\right)+3c_{d}\left(-2m_{\eta}^{2}+t\right)\right)}{9f_{\pi}^{4}\left(m_{S_8}^{2}-t\right)}\nonumber\\
    && -\frac{\left(c_{m}\left(-5m_{K}^{2}+3m_{\pi}^{2}\right)+c_{d}\left(m_{\eta}^{2}+m_{K}^{2}-u\right)\right)^{2}}{6f_{\pi}^{4}\left(m_{S_8}^{2}-u\right)}\nonumber \\
    && -\frac{4\left(2(-\tilde{c}_{d}+\tilde{c}_{m})m_{K}^{2}+\tilde{c}_{d}t\right)\left(-6\tilde{c}_{d}m_{\eta}^{2}+8\tilde{c}_{m}m_{K}^{2}-2\tilde{c}_{m}m_{\pi}^{2}+3\tilde{c}_{d}t\right)}{3f_{\pi}^{4}\left(m_{S_0}^{2}-t\right)}\\
    && +\frac{4d_{m}^{2}\left(-4m_{K}^{4}+3m_{\pi}^{2}m_{K}^{2}\right)}{3f_{\pi}^{4}\left(m_{K}^{2}-m_{P_8}^{2}\right)}+\frac{4d_{m}^{2}\left(-20m_{K}^{4}+9m_{\pi}^{2}m_{K}^{2}-4m_{\pi}^{4}\right)}{9f_{\pi}^{4}\left(m_{\eta}^{2}-m_{P_8}^{2}\right)} -\frac{32\tilde{d}_{m}^{2}\left(2m_{K}^{4}-3m_{\pi}^{2}m_{K}^{2}+m_{\pi}^{4}\right)}{3f_{\pi}^{4}\left(m_{\eta}^{2}-m_{P_0}^{2}\right)}\nonumber\\
    &&+\frac{c_{m}\left(m_{K}^{2}-m_{\pi}^{2}\right)\left(21c_{m}\left(-m_{K}^{2}+m_{\pi}^{2}\right)+c_{d}\left(246m_{K}^{2}-58m_{\pi}^{2}-63(s+u)\right)\right)}{9f_{\pi}^{4}m_{S_8}^{2}}\nonumber.
  \end{eqnarray}

  For the channel $\overline{K}^{0}\eta\to\overline{K}^{0}\pi^{0}$ we have,
  \begin{eqnarray}
    T^{\rm SRA}(s,t,u)&=&-\frac{8m_{K}^{2}+m_{\pi}^{2}+3m_{\eta}^{2}-9t}{12\sqrt{3}f_{\pi}^{2}}-
    \frac{\sqrt{3}G_{V}^{2}\left\{\left(m_{K}^{2}+m_{\pi}^{2}-u\right)\left(m_{K}^{2}+m_{\eta}^{2}-u\right)-\left(2m_{K}^{2}-t\right)\left(m_{\pi}^{2}+m_{\eta}^{2}-t\right)\right\}}{4f_{\pi}^{4}\left(s-m_{V}^{2}\right)}\nonumber \\
    &&-\frac{\sqrt{3}G_{V}^{2}\left\{\left(m_{K}^{2}+m_{\pi}^{2}-s\right)\left(m_{K}^{2}+m_{\eta}^{2}-s\right)-\left(2m_{K}^{2}-t\right)\left(m_{\pi}^{2}+m_{\eta}^{2}-t\right)\right\}}{4f_{\pi}^{4}\left(u-m_{V}^{2}\right)} \nonumber \\
    && +\frac{\left(c_{m}\left(-5m_{K}^{2}+3m_{\pi}^{2}\right)+c_{d}\left(m_{\eta}^{2}+m_{K}^{2}-s\right)\right)\left((c_{d}-c_{m})\left(m_{K}^{2}+m_{\pi}^{2}\right)-c_{d}s\right)}{2\sqrt{3}f_{\pi}^{4}\left(-m_{S_8}^{2}+s\right)}\nonumber \\
    && +\frac{\left(-2c_{m}m_{\pi}^{2}+c_{d}\left(m_{\eta}^{2}+m_{\pi}^{2}-t\right)\right)\left(2(c_{d}-c_{m})m_{K}^{2}-c_{d}t\right)}{\sqrt{3}f_{\pi}^{4}\left(m_{S_8}^{2}-t\right)}\\
    && +\frac{\left(c_{m}\left(-5m_{K}^{2}+3m_{\pi}^{2}\right)+c_{d}\left(m_{\eta}^{2}+m_{K}^{2}-u\right)\right)\left((c_{d}-c_{m})\left(m_{K}^{2}+m_{\pi}^{2}\right)-c_{d}u\right)}{2\sqrt{3}f_{\pi}^{4}\left(u-m_{S_8}^{2}\right)}\nonumber \\
    && +\frac{16\tilde{d}_{m}^{2}\left(-2m_{K}^{4}+m_{K}^{2}m_{\pi}^{2}+m_{\pi}^{4}\right)}{3\sqrt{3}f_{\pi}^{4}\left(m_{\eta}^{2}-m_{P_0}^{2}\right)}+\frac{4d_{m}^{2}m_{K}^{2}\left(-2m_{K}^{2}+m_{\pi}^{2}\right)}{\sqrt{3}f_{\pi}^{4}\left(m_{K}^{2}-m_{P_8}^{2}\right)}\nonumber \\
    && +\frac{2d_{m}^{2}\left(4m_{K}^{4}-5m_{K}^{2}m_{\pi}^{2}+m_{\pi}^{4}\right)}{9\sqrt{3}f_{\pi}^{4}\left(m_{\eta}^{2}-m_{P_8}^{2}\right)}+\frac{2d_{m}^{2}m_{\pi}^{2}\left(-3m_{K}^{2}+m_{\pi}^{2}\right)}{3\sqrt{3}f_{\pi}^{4}\left(m_{\pi}^{2}-m_{P_8}^{2}\right)}\nonumber\\
    &&+\frac{c_{m}\left(m_{K}^{2}-m_{\pi}^{2}\right)\left(c_{m}\left(-m_{K}^{2}+m_{\pi}^{2}\right)+c_{d}\left(8m_{K}^{2}-8m_{\pi}^{2}+4s-11t+4u\right)\right)}{3\sqrt{3}f_{\pi}^{4}m_{S_8}^{2}}\nonumber.
  \end{eqnarray}

  In the case of the process $\pi^{0}\eta\to\pi^{0}\eta$ the SRA amplitude is given by,
  \begin{eqnarray}
    T^{\rm SRA}(s,t,u)&=&-\frac{1}{3f_{\pi}^{2}}m_{\pi}^{2}+\frac{2\left(-2c_{m}m_{\pi}^{2}+c_{d}\left(m_{\eta}^{2}+m_{\pi}^{2}-s\right)\right)^{2}}{3f_{\pi}^{4}\left(s-m_{S_8}^{2}\right)}\nonumber \\
    && +\frac{4\left(2(-\tilde{c}_{d}+\tilde{c}_{m})m_{\pi}^{2}+\tilde{c}_{d}t\right)\left(-6\tilde{c}_{d}m_{\eta}^{2}+8\tilde{c}_{m}m_{K}^{2}-2\tilde{c}_{m}m_{\pi}^{2}+3\tilde{c}_{d}t\right)}{3f_{\pi}^{4}\left(t-m_{S_0}^{2}\right)}\nonumber \\
    && -\frac{2\left(2(-c_{d}+c_{m})m_{\pi}^{2}+c_{d}t\right)\left(2c_{m}\left(8m_{K}^{2}-5m_{\pi}^{2}\right)+3c_{d}\left(-2m_{\eta}^{2}+t\right)\right)}{9f_{\pi}^{4}\left(t-m_{S_8}^{2}\right)} \\
    && +\frac{2\left(-2c_{m}m_{\pi}^{2}+c_{d}\left(m_{\eta}^{2}+m_{\pi}^{2}-u\right)\right)^{2}}{3f_{\pi}^{4}\left(u-m_{S_8}^{2}\right)}-\frac{8d_{m}^{2}m_{\pi}^{4}}{3f_{\pi}^{4}\left(m_{\pi}^{2}-m_{P_8}^{2}\right)}+\frac{16d_{m}^{2}m_{\pi}^{2}\left(m_{\pi}^{2}-4m_{K}^{2}\right)}{9f_{\pi}^{4}\left(m_{\eta}^{2}-m_{P_8}^{2}\right)}\nonumber\\
    &&+\frac{64\tilde{d}_{m}^{2}m_{\pi}^{2}\left(m_{K}^{2}-m_{\pi}^{2}\right)}{3f_{\pi}^{4}\left(m_{\eta}^{2}-m_{P_0}^{2}\right)}+\frac{32c_{d}c_{m}m_{\pi}^{2}(m_{K}^{2}+m_{\pi}^{2})}{9f_{\pi}^{4}m_{S_8}^{2}}\nonumber.
  \end{eqnarray}

  Finally, for the channel $\eta\eta\to\eta\eta$ it reads,
  \begin{eqnarray}
    T^{\rm SRA}(s,t,u)&=&\frac{7m_{\pi}^{2}-16m_{K}^{2}}{9f_{\pi}^{2}}+\frac{4\left(6\tilde{c}_{t}m_{\eta}^{2}-8\tilde{c}_{m}m_{K}^{2}+2\tilde{c}_{m}m_{\pi}^{2}-3\tilde{c}_{d}s\right)^{2}}{9f_{\pi}^{4}\left(s-m_{S_0}^{2}\right)}+\frac{2\left(6c_{d}m_{\eta}^{2}-16c_{m}m_{K}^{2}+10c_{m}m_{\pi}^{2}-3c_{d}s\right)^{2}}{27f_{\pi}^{4}\left(s-m_{S_8}^{2}\right)}\nonumber \\
    && +\frac{4\left(6\tilde{c}_{t}m_{\eta}^{2}-8\tilde{c}_{m}m_{K}^{2}+2\tilde{c}_{m}m_{\pi}^{2}-3\tilde{c}_{d}t\right)^{2}}{9f_{\pi}^{4}\left(t-m_{S_{S_0}}^{2}\right)}+\frac{2\left(6c_{d}m_{\eta}^{2}-16c_{m}m_{K}^{2}+10c_{m}m_{\pi}^{2}-3c_{d}t\right)^{2}}{27f_{\pi}^{4}\left(t-m_{S_8}^{2}\right)}\nonumber\\
    && +\frac{4\left(6\tilde{c}_{t}m_{\eta}^{2}-8\tilde{c}_{m}m_{K}^{2}+2\tilde{c}_{m}m_{\pi}^{2}-3\tilde{c}_{d}u\right)^{2}}{9f_{\pi}^{4}\left(u-m_{S_0}^{2}\right)}+\frac{2\left(6c_{d}m_{\eta}^{2}-16c_{m}m_{K}^{2}+10c_{m}m_{\pi}^{2}-3c_{d}u\right)^{2}}{27f_{\pi}^{4}\left(u-m_{S_8}^{2}\right)} \\
    && -\frac{128\tilde{d}_{m}^{2}\left(5m_{\pi}^{4}+8m_{K}^{4}-13m_{K}^{2}m_{\pi}^{2}\right)}{9f_{\pi}^{4}\left(m_{\eta}^{2}-m_{P_0}^{2}\right)}-\frac{32d_{m}^{2}\left(64m_{K}^{4}-44m_{K}^{2}m_{\pi}^{2}+7m_{\pi}^{4}\right)}{27f_{\pi}^{4}\left(m_{\eta}^{2}-m_{P_8}^{2}\right)}+\frac{128d_{m}^{2}m_{K}^{4}}{9f_{\pi}^{4}\left(m_{K}^{2}-m_{P_8}^{2}\right)}\nonumber\\
    &&-\frac{56d_{m}^{2}m_{\pi}^{4}}{9f_{\pi}^{4}\left(m_{\pi}^{2}-m_{P_8}^{2}\right)}+\frac{64c_{m}\left(m_{K}^{2}-m_{\pi}^{2}\right)\left(10c_{d}m_{K}^{2}-6c_{m}m_{K}^{2}-7c_{d}m_{\pi}^{2}+6c_{m}m_{\pi}^{2}\right)}{27f_{\pi}^{4}m_{S_8}^{2}}\nonumber.
  \end{eqnarray}

  Finally, for the sake of completeness, we have checked for each of the considered processes,
  that the above formulas are equivalent to the $\mathcal{O}\left(p^{4}\right)$ contributions proportional
  to the LECs $L_{i}$ of Ref.~\cite{GomezNicola:2001as}, when the following relations hold:
  \begin{eqnarray}
    && L_{1}=\frac{G_{V}^{2}}{8m_{V}^{2}}-\frac{c_{d}^{2}}{6m_{S_8}^{2}}+\frac{\tilde{c}_{d}^{2}}{2m_{S_0}^{2}},\,\, L_{2}=\frac{G_{V}^{2}}{4m_{V}^{2}},\,\, L_{3}=\frac{-3G_{V}^{2}}{4m_{V}^{2}}+\frac{c_{d}^{2}}{2m_{S_8}^{2}},\nonumber\\
    && L_{4}=-\frac{c_{d}c_{m}}{3m_{S_8}^{2}}+\frac{\tilde{c}_{d}\tilde{c}_{m}}{m_{S_0}^{2}},\,\, L_{5}=\frac{c_{d}c_{m}}{m_{S_8}^{2}},\,\, L_{6}=-\frac{c_{m}^{2}}{6m_{S_8}^{2}}+\frac{\tilde{c}_{m}^{2}}{2m_{S_0}^{2}},\,\,L_{8}=-3L_{6},\\
    && L_{7}=\frac{d_{m}^{2}}{6m_{P_8}^{2}}-\frac{\tilde{d}_{m}^{2}}{2m_{P_0}^{2}},\,\,\,\, L_{8}=\frac{c_{m}^{2}}{2m_{S_8}^{2}}-\frac{d_{m}^{2}}{2m_{P_8}^{2}}.\nonumber
  \end{eqnarray}
\end{widetext}

\bibliography{refs}

\end{document}